\documentclass[preprint2]{aastex}

\usepackage{graphicx}
\usepackage{color}

\shorttitle{The isotropic IPD and EBL}

\shortauthors{Sano et al.}

\begin{document}


\title{The isotropic interplanetary dust cloud and near-infrared extragalactic background light observed with {\it COBE}/DIRBE}


\author{K. SANO\altaffilmark{1,2},
 S. MATSUURA\altaffilmark{2},
 K. YOMO\altaffilmark{2,3},
 A. TAKAHASHI\altaffilmark{4}}

\affil{
\altaffilmark{1}College of Science and Engineering, School of Mathematics and  Physics, Kanazawa University, Kakuma, Kanazawa, Ishikawa 920-1192, Japan\\
\altaffilmark{2}Department of Physics, School of Science and Engineering, Kwansei Gakuin University, 2-1 Gakuen, Sanda, Hyogo 669-1337, Japan \\
\altaffilmark{3}Department of Aerospace Engineering, Tohoku University, 2-1-1, Katahira, Aoba-ku, Sendai, Miyagi, 980-8577, Japan \\
\altaffilmark{4}Astrobiology Center, 2-21-1, Osawa, Mitaka, Tokyo 181-8588, Japan \\
}

\email{sano1989@se.kanazawa-u.ac.jp}






\begin{abstract}
We report observation of isotropic interplanetary dust (IPD) by analyzing the infrared (IR) maps of Diffuse Infrared Background Experiment (DIRBE) onboard the {\it Cosmic Background Explorer} ({\it COBE}) spacecraft.
To search for the isotropic IPD, we perform new analysis in terms of solar elongation angle ($\epsilon$), because we expect zodiacal light (ZL) intensity from the isotropic IPD to decrease as a function of $\epsilon$.
We use the DIRBE weekly-averaged maps covering $64^\circ \lesssim \epsilon \lesssim 124^\circ$ and inspect the $\epsilon$-dependence of residual intensity after subtracting conventional ZL components.
We find the $\epsilon$-dependence of the residuals, indicating the presence of the isotropic IPD.
However, the mid-IR $\epsilon$-dependence is different from that of the isotropic IPD model at $\epsilon \gtrsim 90^\circ$, where the residual intensity increases as a function of $\epsilon$.
To explain the observed $\epsilon$-dependence, we assume a spheroidal IPD cloud showing higher density further away from the sun.
We estimate intensity of the near-IR extragalactic background light (EBL) by subtracting the spheroidal component, assuming the spectral energy distribution from the residual brightness at $12\,{\rm \mu m}$.
The EBL intensity is derived as $45_{-8}^{+11}$, $21_{-4}^{+3}$, and $15\pm3\,{\rm nWm^{-2}sr^{-1}}$ at $1.25$, $2.2$, and $3.5\,{\rm \mu m}$, respectively. 
The EBL is still a few times larger than integrated light of normal galaxies, suggesting existence of unaccounted extragalactic sources.
\end{abstract}





\section{INTRODUCTION}

Interplanetary dust (IPD) is one constituent in our solar system as well as the sun or planets and exists in interplanetary space ubiquitously.
The IPD properties including size distribution or composition have been investigated by in-situ flux measurements of the IPD grains, such as the missions of {\it Helios}, {\it Ulysses}, {\it Galileo}, {\it Cassini}, and {\it New Horizons} (e.g., Hillier et al. 2007; Poppe et al. 2010; Poppe et al. 2011; Szalay et al. 2013).
The IPD properties can also be studied by observations of zodiacal light (ZL), the scattered sunlight or thermal emission from the IPD.
According to observations with a number of ground-based or space telescopes in ultraviolet (UV), visible, and infrared (IR) wavelengths, optical and physical properties of the IPD have been investigated by analyzing the ZL spectra (e.g., Leinert et al. 1981; Matsuura et al. 1995; Matsumoto et al. 1996; Leinert et al. 1998; Tsumura et al. 2010; Krick et al. 2012; Tsumura et al. 2013a; Ishiguro et al. 2013; Yang \& Ishiguro 2015; Kawara et al. 2017; Takahashi et al. 2019).
In the IR wavelengths, several studies have developed parameterized ZL models including spatial distribution and grain properties of the IPD (e.g., albedo, phase function, temperature) on the basis of all-sky observations, such as {\it Infrared Astronomical Satellite} ({\it IRAS}; Wheelock et al. 1994) and Diffuse Infrared Background Experiment (DIRBE) onboard {\it Cosmic Background Explorer} ({\it COBE}) satellite (Kelsall et al. 1998; Wright 1998).

Due to the Poynting-Robertson drag or solar radiation pressure, the IPD is thought to dissipate within $\sim10^3$--$10^7\,{\rm yr}$ (e.g., Burns et al. 1979; Mann et al. 2006). 
This timescale is much shorter than the history of our solar system, indicating that the  IPD grains produced in the proto-planetary phase do not exist in the present epoch.
Therefore, the IPD grains should have been supplied incessantly by some objects, such as asteroids or comets.
Comets are classified as Jupiter Family comets (JFCs; Levison \& Duncan 1997), Halley-type comets (HTCs), and Oort-Cloud Comets (OCCs; Francis 2005).
In the outer solar system of $\sim50\,{\rm AU}$ from the sun, Edgeworth-Kuiper Belt (EKB) is thought to be the main source of the IPD  (Landgraf et al. 2002). 
Around the earth orbit, the IPD from JFCs is thought to prevail widely in low and high ecliptic latitudes against those from asteroids or OCCs (e.g., Nesvorn\'y et al. 2010; Hahn et al. 2002; Poppe 2016).
The conventional ZL models include this IPD component from JFCs as a smooth cloud, in addition to the dust bands originating from the asteroidal IPD (Reach 1992; Spiesman et al. 1995) and circumsolar ring trapped in the earth orbit (Dermott et al. 1994).
On the other hand, the IPD grains supplied by OCCs are thought to show isotropic density distribution around the sun since the Oort cloud is assumed as a shell-shaped isotropic component in the outer solar system (Oort 1950).
A sign of the isotropic IPD component has been reported by observations with Clementine spacecraft (Hahn et al. 2002),  {\it IRAS} (Nesvorn\'y et al. 2010), and {\it AKARI} (Kondo et al. 2016).
Poppe (2016) predicts spatial density distribution of grains from OCCs according to dynamical simulation of the IPD in our solar system.
These studies consistently expect mass fraction of the OCC grains to be less than $\sim10\%$ of the total IPD.
The conventional ZL model developed by Kelsall et al. (1998) do not include the isotropic IPD component from OCCs because the model is created by fitting to seasonal variation of the observed sky brightness.
The investigation on the absolute amount of the isotropic IPD by the ZL observations is necessary to understand the origins of the IPD comprehensively.

In addition to the astrophysical interest of the IPD, the ZL evaluation is also crucial for measurement of extragalactic background light (EBL) in the visible and IR wavelengths because the ZL component should be removed accurately to measure the EBL.
The EBL is an integral constraint on the energy released by cosmic star formation activity, and can be used to constrain energy releases from particular objects, such as primordial black holes, Population III or Dark Stars (e.g., Bond et al. 1986; Aguirre \& Haiman 2000; Hauser \& Dwek 2001; Yue et al. 2013; Maurer et al. 2012).
Moreover, the EBL observation is important for high-energy astrophysics because the GeV--TeV photons from distant sources (e.g., blazars or $\gamma$-ray bursts) are attenuated by the electron-positron pair creation with the EBL photons (e.g., Stanev \& Franceschini 1998; Dwek \& Krennrich 2005; Dwek et al. 2005b; Aharonian et al. 2006; Mazin \& Raue 2007; Meyer et al. 2012; Franceschini et al. 2008; Abdollahi et al. 2018). 
The degree of attenuation is determined by the intensity and spectral shape of the EBL.

To measure the EBL in the visible and near-IR, space observations with sounding rockets or satellites have been conducted by {\it Cosmic Infrared Background Experiment} ({\it CIBER}), {\it Hubble Space Telescope} ({\it HST}), {\it COBE}/DIRBE, {\it Infrared Telescope in Space} ({\it IRTS}), and {\it AKARI} (e.g., Dwek \& Arendt 1998; Brown et al. 2000; Wright \& Reese 2000; Wright 2001; Levenson et al. 2007; Levenson \& Wright 2008; Matsuura et al. 2017; Kawara et al. 2017; Sano et al. 2015; Tsumura et al. 2013c; Matsumoto et al. 2015). 
In these studies, the residual light derived by subtracting foreground emissions from observed sky brightness is regarded as the EBL.
The ZL, one of the foreground emissions, has been estimated and removed using a parameterized ZL model created by the IR all-sky observations with {\it COBE}/DIRBE.
Figure 1 summarizes current measurements of the EBL intensity from UV to IR wavelengths in comparison to the ZL and integrated galaxy light (IGL).
At $0.2$--$4.0\,{\rm \mu m}$, some of them report residual light several times larger than the IGL obtained from deep galaxy counts (Madau \& Pozzetti 2000; Totani et al. 2001; Fazio et al. 2004; Gardner et al. 2000; Xu et al. 2005; Driver et al. 2016).
One explanation for this excess is presence of potential extragalactic objects other than normal galaxies, such as intra halo light (IHL; Cooray et al. 2012b) or direct collapse black holes (DCBH; Yue et al. 2013).
However, most of the EBL constraints from the $\gamma$-ray observations suggest low EBL intensity, comparable to the IGL level. 
Therefore, possibility of ZL underestimation has been discussed as a cause of the excess (e.g., Mattila 2006).
Dwek et al. (2005a) and Kawara et al. (2017) note spectral similarity between the ZL and residuals, indicating presence of the isotropic IPD component that is not included in the conventional ZL model.
To measure the EBL intensity accurately, we now need to investigate contribution of the potential isotropic IPD. 

\begin{figure*}
\begin{center}
 \includegraphics[scale=0.8]{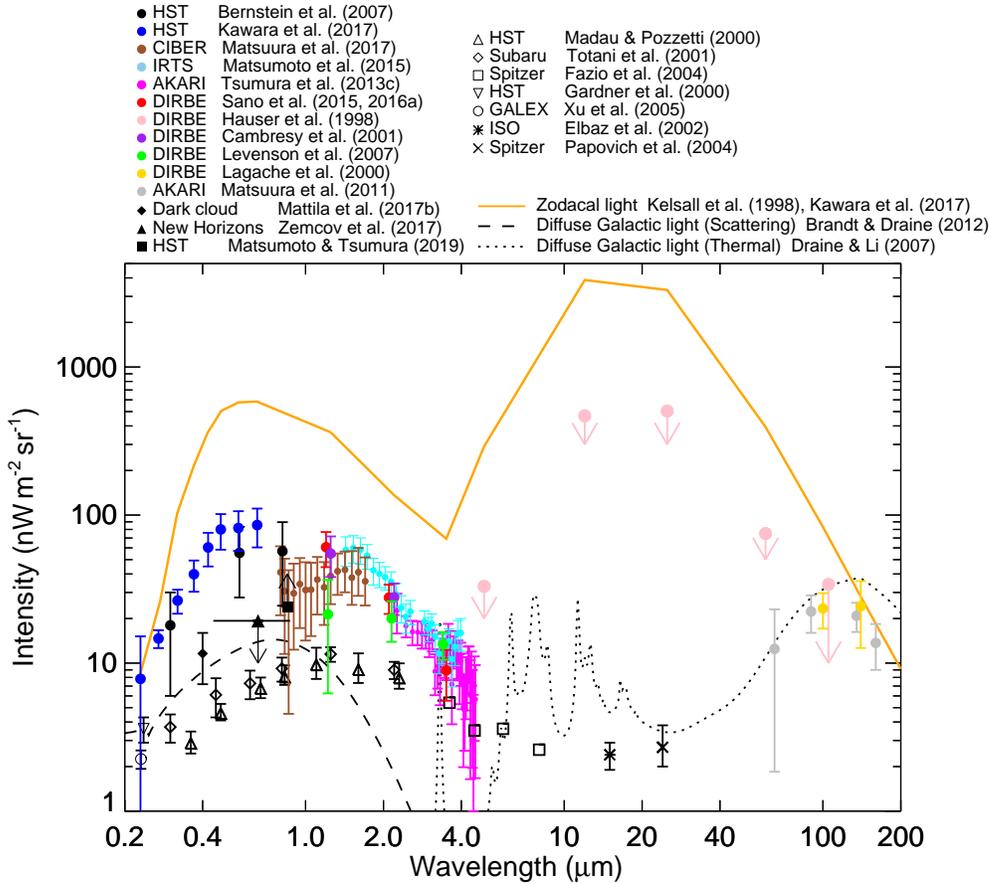}
 \caption
 {Compilation of previous EBL intensity measurements from UV to far-IR in comparison to the IGL and foregrounds, ZL and DGL. 
 Filled symbols represent residual light obtained by subtracting the foreground emissions from the observed sky brightness.
The results from {\it HST}, {\it CIBER}, {\it IRTS}, {\it AKARI}, and {\it DIRBE} are indicated by the filled circles (Bernstein et al. 2007, Kawara et al. 2017, Matsuura et al. 2017, Matsumoto et al. 2015, Tsumura et al. 2013c, Sano et al. 2015, Sano et al. 2016a, Hauser et al. 1998, Cambr\'esy et al. 2001, Levenson et al. 2007, Lagache et al. 2000, and Matsuura et al. 2011).
The filled diamond represents the $0.4\,{\rm \mu m}$ EBL intensity derived by the dark cloud method (Mattila et al. 2017b).
The black triangle indicates upper limit of the visible EBL derived from observations with the {\it New Horizons} spacecraft outer solar system (Zemcov et al. 2017).
The black square shows lower limit of the EBL estimated from power-spectrum analysis of  {\it HST} XDF images (Matsumoto \&  Tsumura 2019).
 Open symbols indicate the IGL intensity derived from deep number counts of galaxies.
 The results from {\it HST}, Subaru, {\it Spitzer}, {\it GALEX}, and {\it ISO} come from Madau \& Pozzetti (2000), Totani et al. (2001), Fazio et al. (2004), Gardner et al. (2000), Xu et al. (2005), Elbaz et al. (2002), and Papovich et al. (2004).
 The near to far-IR orange line denotes the ZL model intensity derived from Kelsall et al. (1998) at  intermediate ecliptic latitudes $\beta \sim 46^\circ$, along with the UV to optical extension according to spectral observation with {\it HST} (Kawara et al. 2017). 
 The dashed and dotted curves are models of scattering and thermal components of the DGL, respectively, according to Brandt \& Draine (2012) and Draine \&  Li (2007).
 The models assume an interstellar radiation field from Mathis et al. (1983) and interstellar dust model from Weingartner \& Draine (2001).
 These spectra are scaled to $1\,{\rm MJy\,sr^{-1}}$ at $100\,{\rm \mu m}$, corresponding to typical diffuse interstellar medium at high latitudes (Schlegel et al. 1998).
}
\end{center}
\end{figure*}

In this paper, we present unprecedented approach to search for the isotropic IPD in the IR wavelengths.
To evaluate the isotropic IPD component, we focus on intensity variation as a function of solar elongation angle ($\epsilon$), which has not been investigated well so far.
In Section 2, we review the conventional DIRBE ZL model and expect $\epsilon$-dependence of the ZL intensity from the isotropic IPD cloud, assuming a simple model of that component.
Section 3 describes all-sky maps created by {\it COBE}/DIRBE observations, covering wide $\epsilon$ range of $64^\circ \lesssim \epsilon \lesssim 124^\circ$.
In Section 4, we show analysis of the DIRBE maps to derive the residual intensity as a function of $\epsilon$. 
The results are also presented in comparison with the isotropic IPD model.
In Section 5, we discuss possible causes of difference between the observed $\epsilon$-dependence and the model prediction.  
Section 6 describes how to separate the near-IR EBL from the isotropic IPD component by using the observed $\epsilon$-dependence of the  residuals.
Section 7 presents implication of the derived EBL in the near-IR in comparison to the IGL, potential extragalactic objects, EBL anisotropy, and $\gamma$-ray constraints. 
Summary of this paper appears in Section 8.

\section{MODELS OF ZODIACAL LIGHT}

\subsection{The DIRBE ZL model}

To introduce an idea of a parameterized ZL model, we briefly review the model according to the all-sky observations with {\it COBE}/DIRBE (Kelsall et al. 1998), hereafter referred to as the Kelsall model.
For $10$ months, DIRBE conducted all-sky observations in $10$ photometric bands at $1.25$, $2.2$, $3.5$, $4.9$, $12$, $25$, $60$, $100$, $140$, and $240\,{\rm \mu m}$ (Hauser et al. 1998).
According to the DIRBE observations, they created all-sky maps with absolute brightness calibration for intensity measurements of diffuse radiation.
To quantify the ZL contribution in the IR wavelengths, the Kelsall model adopt a parameterized physical model, including three-dimensional IPD density distribution and physical properties of the IPD. 
To determine the physical parameters, DIRBE weekly-averaged maps are used to fit the ZL model intensity $I_{\lambda}(p,t)$ at wavelength $\lambda$, a DIRBE pixel $p$, and observation time $t$,
$$
I_{\lambda}(p,t) = \sum_{c} \int n_c(X,Y,Z)[A_{c,\lambda}\,F_{\lambda}^{\odot}\,\phi_{\lambda}(\theta) +
$$
\begin{equation}
 (1-A_{c,\lambda})\,E_{c,\lambda}\,B_{\lambda}(T)\,K_{\lambda}(T)] ds.
\end{equation}
In this formula, $n_c(X,Y,Z)$ is the three-dimensional density distribution of each IPD component $c$, smooth cloud, dust bands, and circumsolar ring in the heliocentric coordinate system $(X,Y,Z)$.
Figure 2(a) shows an IPD grain at heliocentric distance $R$, solar elongation angle $\epsilon$, and ecliptic latitude $\beta$ on the heliocentric coordinates. 
In the Kelsall model, the density distribution of the smooth cloud is assumed to be separable into radial and vertical terms (e.g., Giese et al. 1986),
\begin{equation}
n_c(X,Y,Z) = n_0 R_c^{-\alpha} f(|Z_c/R_c|), 
\end{equation}
where $n_0$, $R_c$ and $Z_c$ denote, respectively, the IPD density at $R=1\,{\rm AU}$, radial and vertical distance from a symmetric plane of the smooth cloud.  
The parameter $\alpha$ is called radial power-law exponent, which is expected to be unity according to the theory of Poynting-Robertson drag (e.g., Gor'kavyi et al. 1997).
The function $f(|Z_c/R_c|)$ is a widened, modified fan model representing exponential and gaussian density distribution toward the vertical direction $Z_c$.
See Kelsall et al. (1998) for more details on the functional forms of the density distribution of the dust bands and circumsolar ring.

\begin{figure*}
\begin{center}
 \includegraphics[scale=0.5,angle=90]{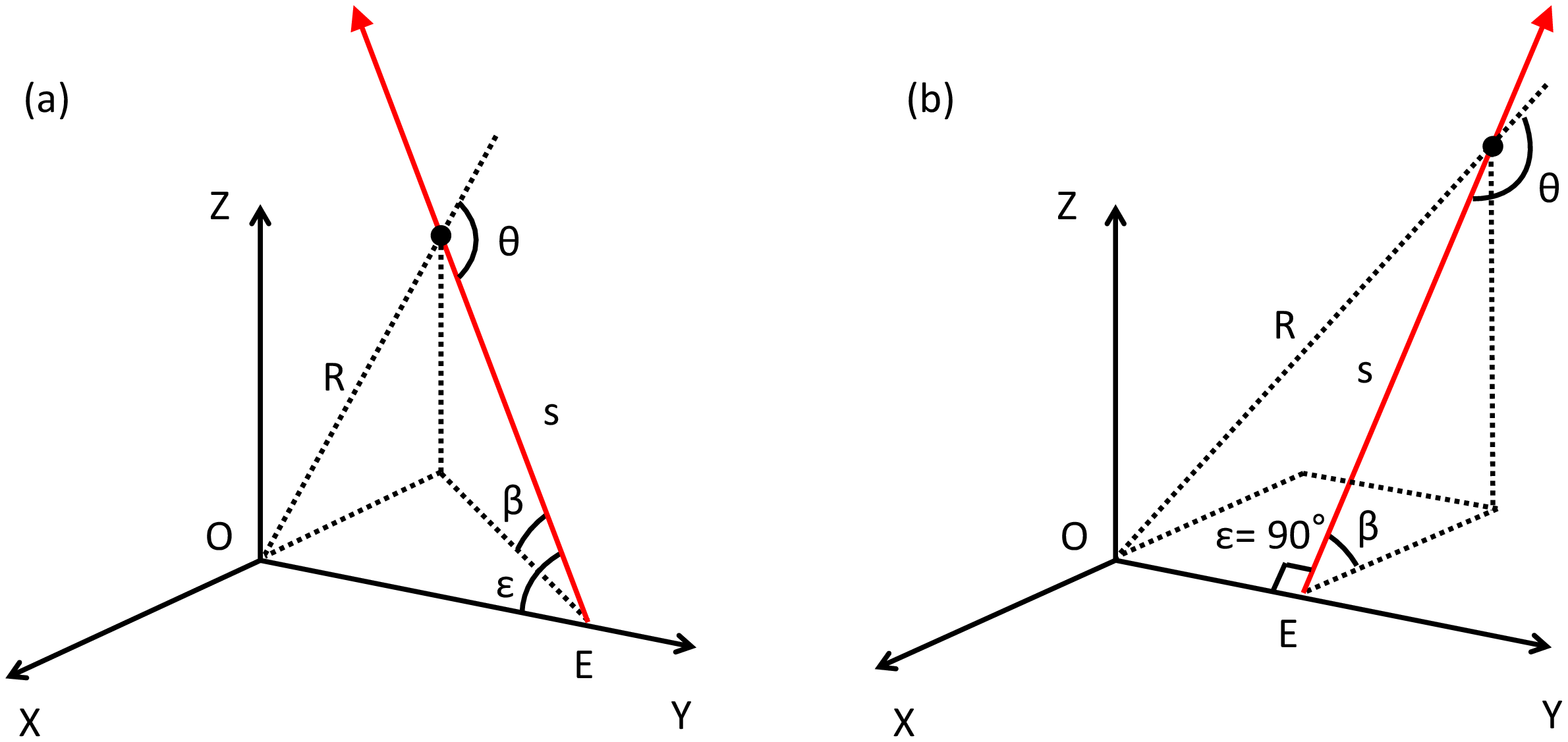}
 \vspace{-30mm}
 \caption
 {Geometry of an IPD grain at the heliocentric distance $R$ in the heliocentric coordinate system $(X,Y,Z)$.
 Positions of the sun and the earth are indicated by ``O'' and ``E'', respectively.
  The red arrow represents line of sight from the earth toward the grain with $s$ indicating distance between the earth and grain.
 Solar elongation angle $\epsilon$ and geocentric ecliptic latitude $\beta$ indicate direction of the grain from the earth.
 Scattering angle toward the earth direction is denoted by $\theta$. 
Panel (a) represents a general situation of the IPD grain, while Panel (b) shows special situation with $\epsilon = 90^\circ$.
Geocentric ecliptic longitude is not shown explicitly.
}
\end{center}
\end{figure*}

In Equation (1), the first and second terms represent scattered light and thermal emission components, respectively.
These elements at position $s$ are integrated from the earth position toward the line of sight to calculate the ZL intensity (Fig. 2).
In the Kelsall model, default integration range is up to $5.2\,{\rm AU}$, close to the orbit radius of Jupiter.
Solar flux at the grain position $R$ is expressed as $F_{\lambda}^{\odot} = F_{\lambda}^{\odot} (R_{\rm E}) / R^2$, where $R_{\rm E}$ is distance between the earth and Sun.
The scattered light intensity is then characterized by grain albedo $A_{c,\lambda}$ and scattering phase function $\phi_{\lambda}(\theta)$ with scattering angle $\theta$.
The phase function is assumed as three-parameter ($C_{0,\lambda}$, $C_{1,\lambda}$, and $C_{2,\lambda}$) functional form reproducing the study of Hong (1985), which is based on a classical form of Henyey \& Greenstein (1941),
\begin{equation}
\phi_{\lambda}(\theta) = N[C_{0,\lambda} + C_{1,\lambda}\, \theta + \exp(C_{2,\lambda}\, \theta)],
\end{equation}
where $N$ is a scaling factor to set integration of this function toward the entire solid angle $4\pi$ to be unity.
On the other hand, the thermal emission component is expressed as emissivity modification factor $E_{c,\lambda}$, Planck function $B_{\lambda}(T)$, and color correction factor for the DIRBE photometric bands $K_{\lambda}(T)$ with grain temperature $T$,
\begin{equation}
T=T_0 R^{-\delta}, 
\end{equation}
where $T_0$ is grain temperature at $R=1\,{\rm AU}$ and  $\delta$ is temperature power-law exponent expected to be $0.5$ for large gray dust in thermal equilibrium.
Representative parameters of the smooth cloud in the Kelsall model are listed in Table 1.
Since the IPD is thought to show wide-range size distribution from micrometer-sized dust to meteors (e.g., Gr\"un et al. 1985; Dikarev \& Gr\"un 2002) and spatial dependence of the grain properties (e.g., Lumme \& Bowell 1985; Renard et al. 1995), the physical parameters derived in the Kelsall model represent averaged properties of the IPD.
By implementing the integration in Equation (1), the ZL brightness can be calculated as functions of the ecliptic coordinates and observation time.
The IDL code to calculate the ZL intensity is available from the DIRBE  website ``lambda.gsfc.nasa.gov/product/cobe/''.

\begin{table*}
\begin{center}
 \renewcommand{\arraystretch}{1.0}
 \caption{Some IPD parameters of the smooth cloud in the Kelsall model}
  \label{symbols}
  \begin{tabular}{llcc}
  \hline
   Parameter & Description & Final value &  Uncertainty \\
 \hline
   $n_0 ({\rm AU^{-1}})$  & Density at $1\,{\rm AU}$    & $1.13\times 10^{-7}$ & $6.4\times 10^{-10}$ \\
   $\alpha$  & Radial power-law exponent   & $1.34$ & $0.022$ \\
   $T_0 ({\rm K})$  & Temperature  at $1\,{\rm AU}$ & $286$ & Fixed  \\ 
   $\delta$ &  Temperature power-law exponent   & $0.467$ & $0.0041$\\ 
       \hline
        \end{tabular}
    \end{center}
    \medskip
        
 \end{table*}

In the Kelsall model, seasonal variation of the sky brightness is fitted by the physical model (Equation 1).
Therefore, isotropic or nearly isotropic IPD components supplied from OCCs are canceled out and excluded from the model.
Hauser et al. (1998) note that the model uncertainty from the missing isotropic IPD influences the resultant residual intensity.
The residuals derived by using the Kelsall model probably contain the isotropic IPD component and this can cause overestimation of the EBL.
To evaluate an amount of the isotropic IPD is crucial for the EBL measurement.

\subsection{A ZL model from the isotropic IPD}

\begin{figure*}
\begin{center}
 \includegraphics[scale=0.5,angle=0]{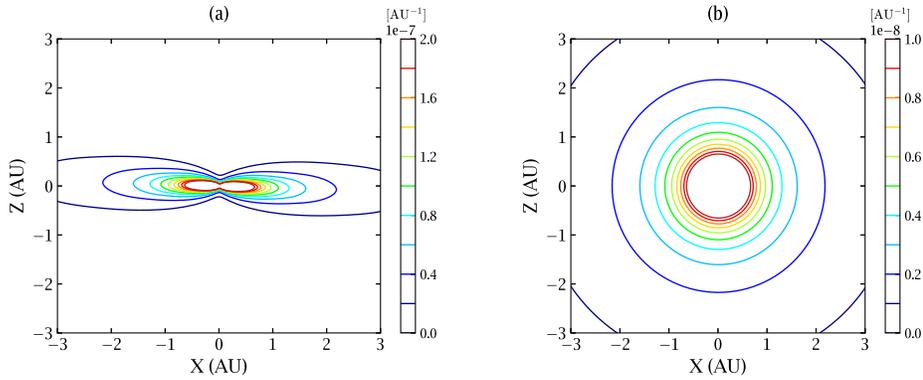}
  \vspace{50mm}
 \caption
 {Cross-sectional density distribution of the smooth cloud in the Kelsall model (a) and the isotropic IPD (b) assuming the same radial density exponent $\alpha=1.34$ (Table 1).
Density of the isotropic IPD is set as $5\%$ of that of the smooth cloud at $1\,{\rm AU}$.
}
\end{center}
\end{figure*}

We examine properties of the isotropic IPD in terms of the intensity dependence on ecliptic latitude ($\beta$) or solar elongation ($\epsilon$).
In general, $\beta$-dependence is used as a measure of intensity variation of the ZL. 
As a specific case of Fig. 2(a), Fig. 2(b) illustrates a situation of solar elongation angle $\epsilon=90^\circ$. 
This situation is applied to some previous satellite observations whose observable regions are limited to $\epsilon\sim90^\circ$ (e.g., {\it AKARI}).
Figure 3 compares density distribution of the smooth cloud in the Kelsall model and that of the isotropic IPD assumed as $n(R) \propto R^{-\alpha}$ with $\alpha=1.34$.
It is obvious from Fig. 2(b) and 3(b) that the ZL intensity from the isotropic IPD does not show $\beta$-dependence in the case of $\epsilon=90^\circ$.
In contrast, we can expect $\epsilon$-dependence of the ZL intensity from the isotropic IPD because the solar flux and dust temperature at a grain position change as a function of $\epsilon$ (Fig. 2a).
This test implies that we should investigate not $\beta$-dependence but $\epsilon$-dependence of the ZL brightness to study properties of the isotropic IPD.   

To examine the $\epsilon$-dependence of the ZL intensity from the isotropic IPD, we calculate it by assuming the density distribution shown in Fig. 3(b).
Nesvorn\'y et al. (2010) compare the mid-IR ${\it IRAS}$ data with dynamical simulation of the IPD grains and constrain the density of the isotropic OCC dust to be less than $\sim 10\%$ of that of the total IPD.
In the Kelsall model, the IPD density of the smooth cloud at $R = 1\,{\rm AU}$ is $n_0 = 1.13\times10^{-7}\,{\rm AU^{-1}}$ (Table 1).
We then adopt $\sim5\%$ of that value, $5.0\times10^{-9}\,{\rm AU^{-1}}$ for the isotropic IPD density.
Other physical parameters of the isotropic IPD is assumed to be same as those determined in the Kelsall model.
According to Fig. 2, geometric parameters $R\,{\rm(AU)}$, $s\,{\rm(AU)}$, and $\theta\,{\rm(rad)}$ are related to $\epsilon$ as   
\begin{equation}
R = \sqrt{s^2-2s\cos\epsilon+1},
\end{equation}
\begin{equation}
\cos\theta = \frac{1-(s^2+R^2)}{2sR}.
\end{equation}
From these formulae and Equation (1), the ZL intensity can be calculated as a function of $\epsilon$.
We adopt the line-of-sight integration toward $50\,{\rm AU}$, motivated by the simulation of the OCC dust density as a function of $R$ (Poppe 2016). 

Figure 4 shows the ZL intensity from the isotropic IPD at $1.25$ and $25\,{\rm \mu m}$ where the scattered light and thermal emission are dominant, respectively (Fig. 1).
For comparison, $\epsilon$-dependence of the IPD components in the Kelsall model is also plotted.
In both wavelengths, the intensity from the isotropic IPD decreases toward high-$\epsilon$ regions in almost all the $\epsilon$.
The scattered light at $1.25\,{\rm \mu m}$ shows a slight turnover in high-$\epsilon$ regions due to backscattering effect in the phase function.
To create the Kelsall model from observations toward various $\epsilon$, {\it COBE}/DIRBE observed wide $\epsilon$ range of $64^\circ \lesssim \epsilon \lesssim 124^\circ$ (shaded regions in Fig. 4).
In these regions, the ZL intensity from the isotropic IPD is expected to decrease simply as a function of $\epsilon$.
These tests indicate that the $\epsilon$-dependence is useful to study the isotropic IPD.
If the isotropic IPD is present, we expect to find the $\epsilon$-dependence of the residual light derived by subtracting other emission components from the total sky brightness.

\begin{figure*}
\begin{center}
 \includegraphics[scale=0.8]{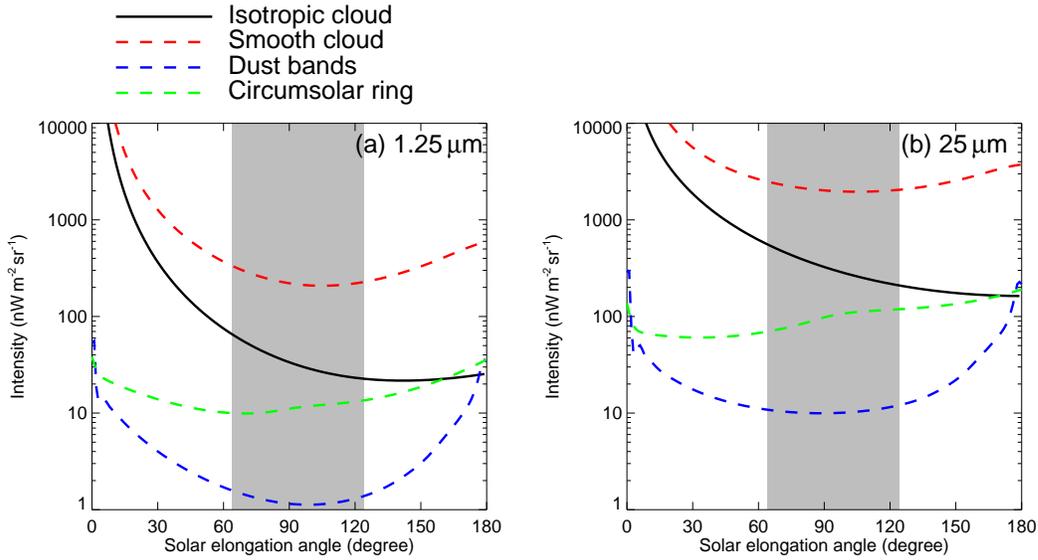}
 \caption
 {Solar-elongation dependence of the intensity of the scattered light and thermal emission from the isotropic IPD at (a) $1.25\,{\rm \mu m}$ and (b) $25\,{\rm \mu m}$, respectively.
Black solid curves represent the prediction from the isotropic IPD component assumed in Section 2.2, while red, blue, and green dashed curves indicate, respectively, the intensity from the smooth cloud, dust bands, and circumsolar ring in the Kelsall model.
The intensity of the IPD components in the Kelsall model is calculated so that the line of sight toward $\epsilon=0^\circ$, $90^\circ$, and $180^\circ$ corresponds to $\beta=0^\circ$, $90^\circ$, and $0^\circ$, respectively.
The shaded region shows nominal coverage of the weekly-averaged maps of DIRBE ($64^\circ \lesssim \epsilon \lesssim 124^\circ$).
}
\end{center}
\end{figure*}

\section{THE DIRBE WEEKLY-AVERAGED MAP}

To investigate the $\epsilon$-dependence as considered in Section 2, we need observations covering wide $\epsilon$ range.
To develop the ZL model, the DIRBE instrument was designed to observe the all sky from  $\epsilon = 64^\circ$ to $124^\circ$ with an optical axis $30^\circ$ offset from a spin axis of the spacecraft. 
One of the DIRBE data products, the weekly-averaged maps, hereafter referred to as weekly maps, were created from daily sky maps in the $10$  photometric bands (COBE/DIRBE Explanatory Supplement 1998).
The data consist of the intensity maps in 41 weeks from Week 4 to 44, during the 10-months cryogenic mission from 1989 November 24 to 1990 September 21.
For scientific analyses,  data reduction and absolute calibration of the available maps are already done by the DIRBE team. 
For example, Fig. 5 shows the weekly map of Week 4 at $1.25\,{\rm \mu m}$.
Panel (a) illustrates the intensity map, while Panel (b) is the corresponding solar elongation angle map covering $64^\circ \lesssim \epsilon \lesssim 124^\circ$.
The $\epsilon$ value at each pixel is calculated as an average during the Week 4 period. 
Therefore, one value of solar elongation angle is assigned to each pixel (Fig. 5b).
The DIRBE data products including the weekly maps are accessible from ``lambda.gsfc.nasa.gov/product/cobe/''.

\begin{figure*}
\begin{center}
 \includegraphics[scale=0.6]{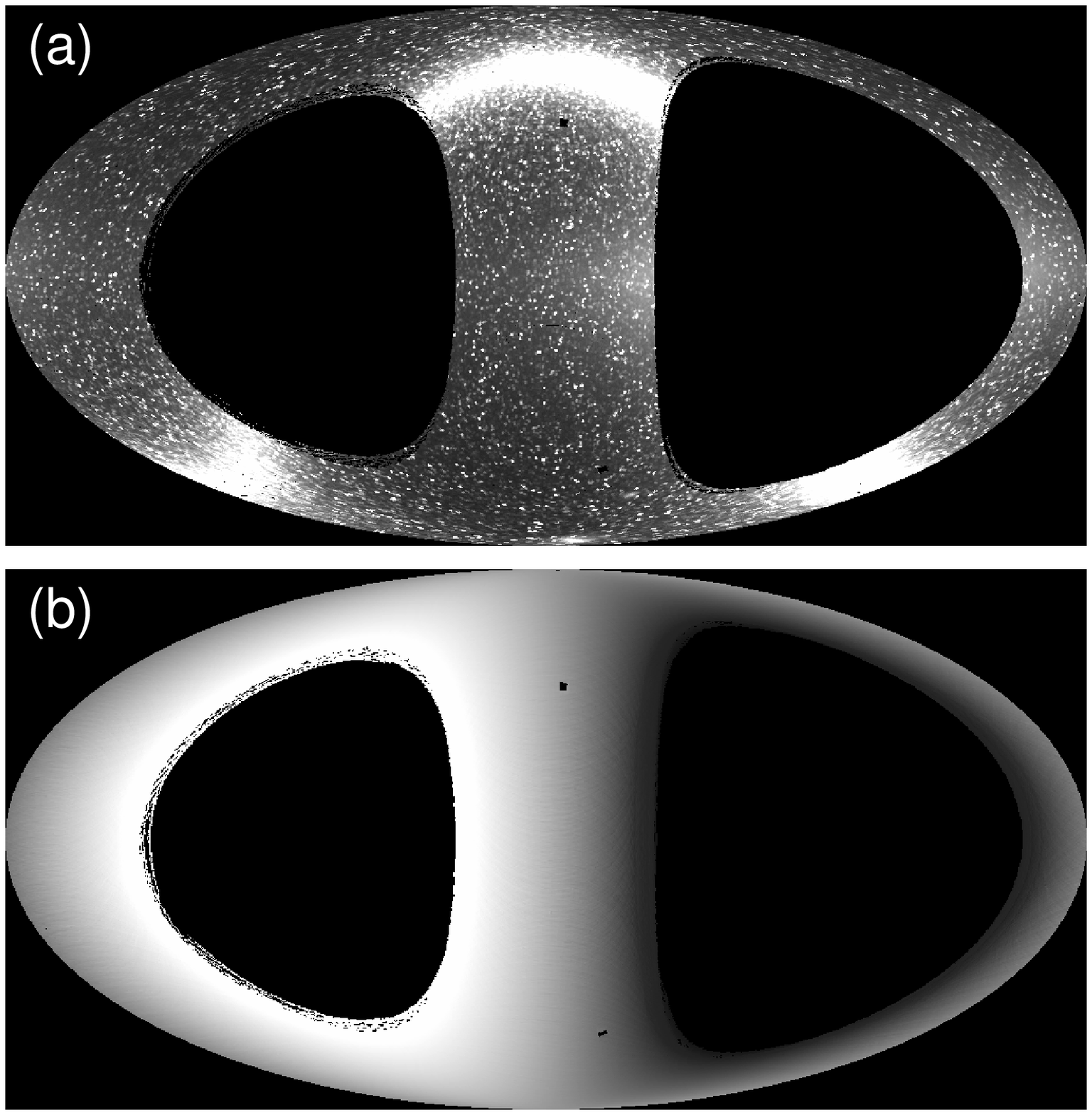}
 \vspace{30mm}
 \caption
 {DIRBE weekly maps of intensity (a) and solar elongation angle (b) at $1.25\,{\rm \mu m}$ in Week 4 in ecliptic Mollweide projection.
 Masked areas correspond to regions out of the DIRBE coverage, i.e., $\epsilon \lesssim 64^\circ$ or $ \epsilon \gtrsim 124^\circ$.
  The ecliptic plane runs horizontally through the center of these images.
 Bright regions in Panel (a) indicates the Galactic plane.
}
\end{center}
\end{figure*}

\section{DERIVATION OF THE RESIDUALS AS A FUNCTION OF SOLAR ELONGATION ANGLE}

We use the weekly maps in the near- and mid-IR at $1.25$, $2.2$, $3.5$, $4.9$, $12$, $25$, and $60\,{\rm \mu m}$.
We do not analyze the data in the longer three DIRBE bands because the ZL intensity is lower and comparable to the EBL level in contrast to the near- and mid-IR situations (Fig. 1).
To avoid fields of a number of Galactic stars and complicated structures of the ZL near the ecliptic plane, analyzed regions are limited to high Galactic and ecliptic latitudes, $|b|>35^\circ$ and $|\beta|>30^\circ$.

\subsection{Near-IR analysis}

In the near-IR bands at $1.25$, $2.2$, $3.5$, and $4.9\,{\rm \mu m}$, astrophysical components are the ZL, diffuse Galactic light (DGL), integrated starlight (ISL), and residual light which includes the EBL and the ZL component from the isotropic IPD. 
In this paper, the DGL means both scattered light and thermal emission from interstellar dust illuminated by interstellar radiation field (e.g., Mathis et al. 1983).
Observed sky intensity $I_{\lambda, i} ({\rm Obs})$ at a wavelength $\lambda$ and a DIRBE pixel $i$ is then assumed as
\begin{equation}
I_{\lambda, i} ({\rm Obs}) = I_{\lambda, i} ({\rm ZL}) + I_{\lambda, i} ({\rm DGL}) + I_{\lambda, i}({\rm ISL}) + I_{\lambda, i} ({\rm Res}),
\end{equation} 
where $I_{\lambda, i} ({\rm ZL})$,  $I_{\lambda, i} ({\rm DGL})$,  $I_{\lambda, i}({\rm ISL})$, and $I_{\lambda, i} ({\rm Res})$ represent intensity of the ZL, DGL, ISL, and residual light, respectively.

Evaluation of the ZL, DGL, ISL is based on our previous papers, Sano et al. (2015, Paper I) and Sano et al. (2016a, Paper II).
In these papers, we assume each component as
\begin{eqnarray}
I_{\lambda, i} ({\rm ZL}) &=& a_{\lambda} I_{\lambda, i} ({\rm Kel}),\\
I_{\lambda, i} ({\rm DGL}) &=& b_{\lambda} I_{100, i},\\
I_{\lambda, i} ({\rm ISL}) &=& c_{\lambda} I_{\lambda, i} ({\rm DISL}),\\
I_{\lambda, i} ({\rm res}) &=& d_{\lambda},
\end{eqnarray}
where $I_{\lambda, i} ({\rm Kel})$, $I_{100, i}$, and $I_{\lambda, i} ({\rm DISL})$ denote, respectively, intensity of the ZL predicted by the Kelsall model, interstellar $100\,{\rm \mu m}$ emission derived by subtracting the EBL component (Lagache et al. 2000) from the {\it IRAS}/DIRBE $100\,{\rm \mu m}$ map (Schlegel et al. 1998), and ISL calculated by integrating fluxes of Galactic stars cataloged by near-IR  all-sky surveys, 2 Micron All-Sky Survey (2MASS; Skrutskie et al. 2006) and {\it Wide-field Infrared Survey Explorer} ({\it WISE}; Wright et al. 2010).
Parameters $a_{\lambda}$, $b_{\lambda}$, $c_{\lambda}$, and $d_{\lambda}$ are free parameters to be determined by fitting the intensity model (Equation 7) to the DIRBE data.
The parameter $a_{\lambda}$ (Equation 8) indicates a correction factor to the Kelsall model and $c_{\lambda}$ (Equation 10) contribution from stars fainter than detection limits of the 2MASS or {\it WISE} catalogs. 
The parameter $b_{\lambda}$ (Equation 9) is motivated by earlier observations that intensity of the visible and near-IR DGL shows linear correlation against that of the $100\,{\rm \mu m}$ emission from interstellar dust (e.g., Ienaka et al. 2013; Tsumura et al. 2013b; Arai et al. 2015; Onishi et al. 2018).
Motivated by theoretical study of Jura (1979), Sano et al. (2016b) and  Sano \& Matsuura (2017) present analysis on $b$-dependence of the parameter $b_{\lambda}$ caused by anisotropic scattering of starlight by interstellar dust and $b$-dependence of the $100\,{\rm \mu m}$ emission.
We take into account the $b$-dependence as one of the uncertainties of the resultant EBL intensity (Section 6.3).

In Paper I and II, the parameters $a_{\lambda}$, $b_{\lambda}$, $c_{\lambda}$, and $d_{\lambda}$ are determined by fitting to one DIRBE product, $\epsilon = 90^\circ$ map created by averaging the weekly maps when each pixel position is close to $\epsilon = 90^\circ$ (COBE/DIRBE Explanatory Supplement 1998).
Thanks to the averaging, the $\epsilon = 90^\circ$ maps show higher signal-to-noise ratios for diffuse radiation than a unit of weekly map and are more suitable to measure the faint DGL component, which is a main motivation of Paper I and II.
Therefore, we adopt the parameters $a_{\lambda}$, $b_{\lambda}$, and $c_{\lambda}$ derived in Paper I and II to calculate the residual intensity $I_{\lambda} (\epsilon)$ in each weekly map, i.e.,  $I_{\lambda} (\epsilon) = I_{\lambda, i} ({\rm Obs})- I_{\lambda, i} ({\rm ZL}) - I_{\lambda, i} ({\rm DGL}) - I_{\lambda, i} ({\rm ISL})$, where $I_{\lambda, i} ({\rm Obs})$ denotes the observed sky brightness in the weekly maps.
Table 2 shows the parameter values of $a_{\lambda}$ and $d_{\lambda}$ with their uncertainties $\sigma(a_\lambda)$ and $\sigma(d_\lambda)$.
The values at $1.25$ and $2.2\,{\rm \mu m}$ are slightly different from those derived in Paper I because the DGL evaluation is set to be same as in Paper II.
The values of $\sigma(a_\lambda)$ include regional variation of the parameter, while those of $\sigma(d_\lambda)$ indicate the statistical uncertainties derived from the fitting (Paper I and II).
Though the $a_\lambda$ values at $1.25$ and $2.2\,{\rm \mu m}$ are close to unity within $\sim5\%$, those at $3.5$ and $4.9\,{\rm \mu m}$ are $\sim 10\%$--$15\%$ larger than unity.
This trend is also reported by Tsumura et al. (2013a) and Matsumoto et al. (2015) in analysis of the {\it AKARI} and {\it IRTS} data, respectively.
At $3.5$ and $4.9\,{\rm \mu m}$, therefore, we also calculate the residuals $I_{\lambda} (\epsilon)$ by assuming $a_{\lambda} = 1.0$ to see systematic effect caused by the deviation from the unity.

\begin{table*}
\begin{center}
 \renewcommand{\arraystretch}{1.0}
 \caption{Parameter values of $a_\lambda$ and $d_\lambda$ and their uncertainties $\sigma(a_\lambda)$ and $\sigma(d_\lambda)$ (Equations 8 and 11)}
  \label{symbols}
  \begin{tabular}{lccccccc}
  \hline
   Band ($\rm{\mu m}$) & $1.25$ & $2.2$ & $3.5$ & $4.9$ & $12$ & $25$ & $60$ \\
 \hline
   $a_\lambda$  (dimensionless)    & $1.008$ & $1.045$ & $1.153$ & $1.100$ & $1.036$ & $1.035$ & $1.016$\\
   $\sigma(a_\lambda)$  (dimensionless)    & $0.012$ & $0.012$ & $0.028$ & $0.051$ & $0.036$ & $0.047$ & $0.068$\\
   $d_\lambda\,({\rm nW\,m^{-2}\,sr^{-1}})$    & $60.66$ & $27.69$ & $8.92$ & $2.67$ & $29.02$ & $62.06$ & $21.10$ \\ 
  $\sigma(d_\lambda)\,({\rm nW\,m^{-2}\,sr^{-1}})$    & $0.08$ & $0.04$ & $0.04$ &  $0.05$ & $0.07$ & $0.06$ & $0.04$ \\ 
       \hline
        \end{tabular}
    \end{center}
    \medskip
 \end{table*}

The uncertainty at each pixel $\sigma_{\lambda, i}$ is calculated as
\begin{equation}
\scalebox{0.9}{$\displaystyle
\sigma_{\lambda, i}^2 = \sigma_{\lambda, i} ({\rm Obs})^2+[\sigma(a_\lambda) I_{\lambda, i} ({\rm ZL})]^2+[\sigma(b_\lambda) I_{100, i}]^2+[\sigma(c_\lambda) I_{\lambda, i} ({\rm ISL})]^2$},
\end{equation}
where $\sigma_{\lambda, i} ({\rm Obs})$, $\sigma(a_\lambda)$, $\sigma(b_\lambda)$, and $\sigma(c_\lambda)$ denote, respectively, uncertainties of the DIRBE weekly map, $a_\lambda$, $b_\lambda$, and $c_\lambda$.
The values $\sigma(b_\lambda)$, and $\sigma(c_\lambda)$ also include the regional variation (Paper I and II).
In the calculation of the $a_{\lambda} = 1.0$ case at $3.5$ and $4.9\,{\rm \mu m}$, $\sigma(a_\lambda)$ is assumed as the same values as shown in Table 2.

\subsection{Mid-IR analysis}

\begin{figure*}
\begin{center}
 \includegraphics[scale=0.8]{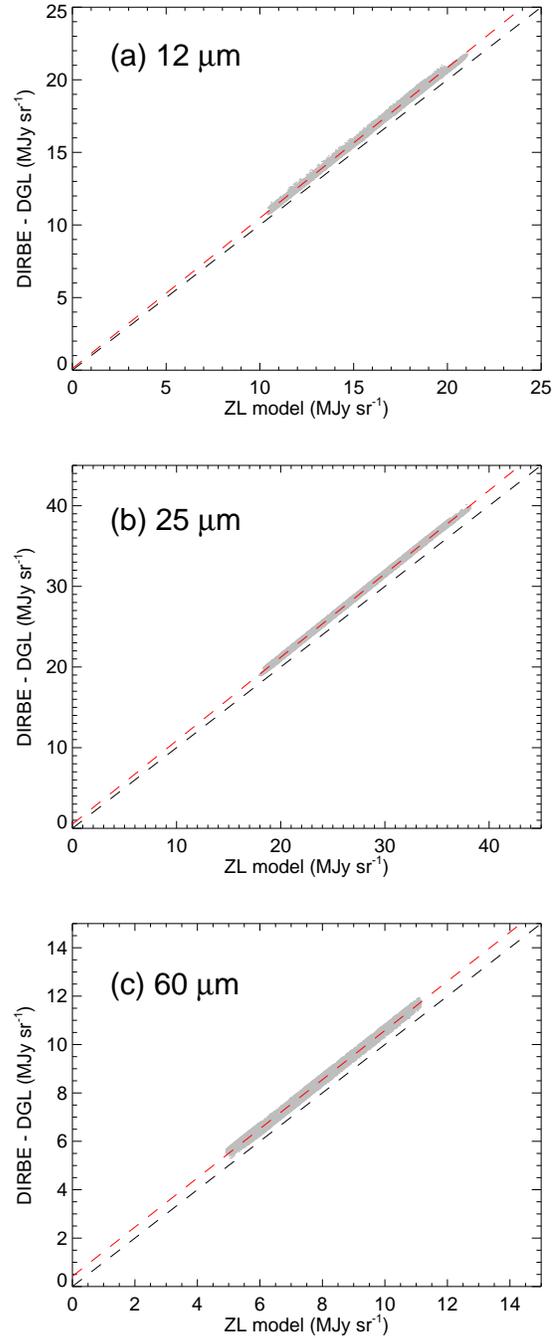}
 \caption
 {Correlation of the intensity of the DGL-subtracted DIRBE $\epsilon = 90^\circ$ map against that of the Kelsall model at (a) $12$, (b) $25$, and (c) $60\,{\rm \mu m}$.
Gray dots indicate the data points.
A red dashed line indicates best-fit line after adopting $2\sigma$ clipping, while a black dashed line denotes a line of $y = x$.
 The best-fit parameters of the lines are shown in Table 2. 
}
\end{center}
\end{figure*}

At $12$, $25$, and $60\,{\rm \mu m}$, the ZL is known to be brighter than the ISL by more than three orders of magnitude (e.g., Fig. 1 of Matsuura et al. 2011).
Therefore, only the DGL and ZL components should be subtracted from the DIRBE weekly maps to derive the residual intensity $I_{\lambda} (\epsilon)$. 
Analyzing the DIRBE data, Arendt et al. (1998) show linear correlations between the mid-IR intensity and $100\,{\rm \mu m}$ emission in high Galactic latitudes.
The derived values of $b_{\lambda}$ are $0.0462\pm0.0001$, $0.0480\pm0.0002$, and $0.171\pm0.0003\,{\rm MJy\,sr^{-1}/MJy\,sr^{-1}}$ at $12$, $25$, and $60\,{\rm \mu m}$, respectively (Table 4 of Arendt et al. 1998).
Adopting these values and Equation (9), we subtract the DGL component from the $\epsilon = 90^\circ$ map at each band.
We then determine the parameters $a_\lambda$ and $d_\lambda$ in the mid-IR by fitting to the $\epsilon = 90^\circ$ map, assuming the DGL-subtracted sky brightness as
\begin{equation}
 I_{\lambda, i} ({\rm Obs}) - I_{\lambda, i} ({\rm DGL}) = a_{\lambda} I_{\lambda, i} ({\rm Kel}) + d_{\lambda}.
\end{equation}
In the fitting, outliers due to bright point sources in the maps are excluded by $2\sigma$ clipping.

The fitting results are shown in Figure 6 and Table 2.
In the three bands, linear correlations are clearly seen with the $a_{\lambda}$ values a few percent larger than unity.
Also, positive values of $d_{\lambda}$ imply presence of the isotropic IPD component that is not included in the Kelsall model.
Then, $\epsilon$-dependence of the residual light can be obtained as $I_{\lambda} ({\epsilon}) = I_{\lambda, i} ({\rm Obs}) - a_{\lambda} I_{\lambda, i} ({\rm Kel}) - b_\lambda I_{100, i}$. 
Since the ZL intensity in the mid-IR is more dominant than that in the near-IR (Fig. 1), the residual intensity should be more sensitive to the $a_{\lambda}$ value.
Therefore, we also calculate the residual intensity by assuming $a_{\lambda} =1.0$ to evaluate difference between the resultant residual intensity.

Uncertainty at each pixel is calculated by Equation (12) with $I_{\lambda, i} ({\rm ISL}) = 0$.
Since the $\sigma(a_\lambda)$ values inferred from the fitting are small due to the high signal-to-noise ratios (Fig. 6), we assume $\sigma(a_\lambda)$ as ratios of nominal uncertainty of the Kelsall model (Table 7 of Kelsall et al. 1998) to the ZL  intensity at a region of intermediate ecliptic latitude (Table 4 of Kelsall et al. 1998).
This makes the $\sigma(a_\lambda)$ values to a few percent of $a_\lambda$  (Table 2).

\subsection{Weekly $\epsilon$-dependence of the residuals}

Figure 7 to 13 represent the residual intensity as a function of $\epsilon$ in each week from $1.25$ to $60\,{\rm \mu m}$.
Individual panels indicate the results from Week 4 to 44.
The residual intensity $\lambda I_{\lambda} ({\epsilon})$ calculated by using the fitting results of $a_{\lambda}$ (Table 2) is represented by black dots, hereafter referred to as Model A.
Except at $1.25$ and $2.2\,{\rm \mu m}$, $\lambda I_{\lambda} ({\epsilon})$ calculated by assuming $a_{\lambda} = 1.0$ is represented by gray dots, hereafter referred to as Model B.
These dots represent weighted-average values of $\lambda I_{\lambda} ({\epsilon})$ within $\Delta\epsilon = 3^\circ$ bins.
Sizes of the dots are proportional to the number of points within the individual bins.
Each bin typically contains $\sim1000$--$2000$ data points.
In several weeks, no data are available around regions of $\epsilon = 90^\circ$.
Such points are excluded from the plots. 
Since junction field effect transistor (JFET) was tested during Week 24, the number of available pixels in that week is significantly lower than the others (COBE/DIRBE Explanatory Supplement 1998).

At $1.25$ and $2.2\,{\rm \mu m}$, most of the weeks exhibit decrease of $\lambda I_{\lambda} (\epsilon)$ as a function of $\epsilon$ in particularly low-$\epsilon$ regions, though the trend changes week by week.
This trend is similar to the $\epsilon$-dependence of the scattered light intensity expected from the isotropic IPD assumed in Section 2.2 (Fig. 4), indicating the presence of the isotropic IPD component.

At $3.5$ and $4.9\,{\rm \mu m}$, the $\epsilon$-dependence is clearly different between Model A and B.
In Model A, $\lambda I_{\lambda} (\epsilon)$ increases toward high-$\epsilon$ regions and several points are negative at $4.9\,{\rm \mu m}$, which are contrary to the $\epsilon$-dependence expected from the isotropic IPD (Fig. 4).
The ZL components may be subtracted excessively in the low-$\epsilon$ regions where the ZL is brighter, due to the $a_{\lambda}$ values larger than 1.0 by $\sim 10\%$ (Table 2).
In contrast, the Model B results show decrease of the intensity toward high-$\epsilon$ regions in most of the weeks, similar to the trends at $1.25$ and $2.2\,{\rm \mu m}$.
Though Model B is close to the trend predicted from the isotropic IPD, we take into account the difference between Model A and B for the final result because the $a_{\lambda}$ value in Model A is the best-fit to the DIRBE data (Paper II).

At $12$ and $25\,{\rm \mu m}$, results from both Model A and B show decrease toward regions around $\epsilon = 90^\circ$ and increase toward the solar elongation extrema, $\sim 64^\circ$ and $\sim124^\circ$. 
This $\epsilon$-dependence is different from the prediction of the isotropic IPD  (Fig. 4).
The residual intensity in Model B is about twice as large as that in Model A because the ZL is much brighter than the residuals in these bands. 
At $60\,{\rm \mu m}$, difference between Model A and B is smaller, though the $\epsilon$-dependence shows significant change through the $41$ weeks. 
This variability might be related to relative faintness of the ZL in this band, though the reason is unclear.

Difference between the observed intensity of DIRBE and the ZL brightness predicted by the Kelsall model is illustrated in Fig. 6 of Kelsall et al. (1998).
The $\epsilon$-dependence found in the present analysis is not evident in their figures probably because averaging individual data points is not applied in their illustrations in contrary to our figures (Fig. 7--13). 

Except at $60\,{\rm \mu m}$, trends of the $\epsilon$-dependence are roughly similar through the $41$ weeks at each wavelength.
Though the plots in Fig. 7 to 13 include only statistical uncertainty of the residuals, it is not reasonable that systematic uncertainty of instrumental origin exhibits the similar $\epsilon$-dependence among the different weeks because correspondence between a DIRBE pixel and $\epsilon$ changes week by week.
Therefore, the observed $\epsilon$-dependence is supposed to come from not instrumental but astrophysical origin.

\begin{figure*}
\begin{center}
 \includegraphics[scale=0.7]{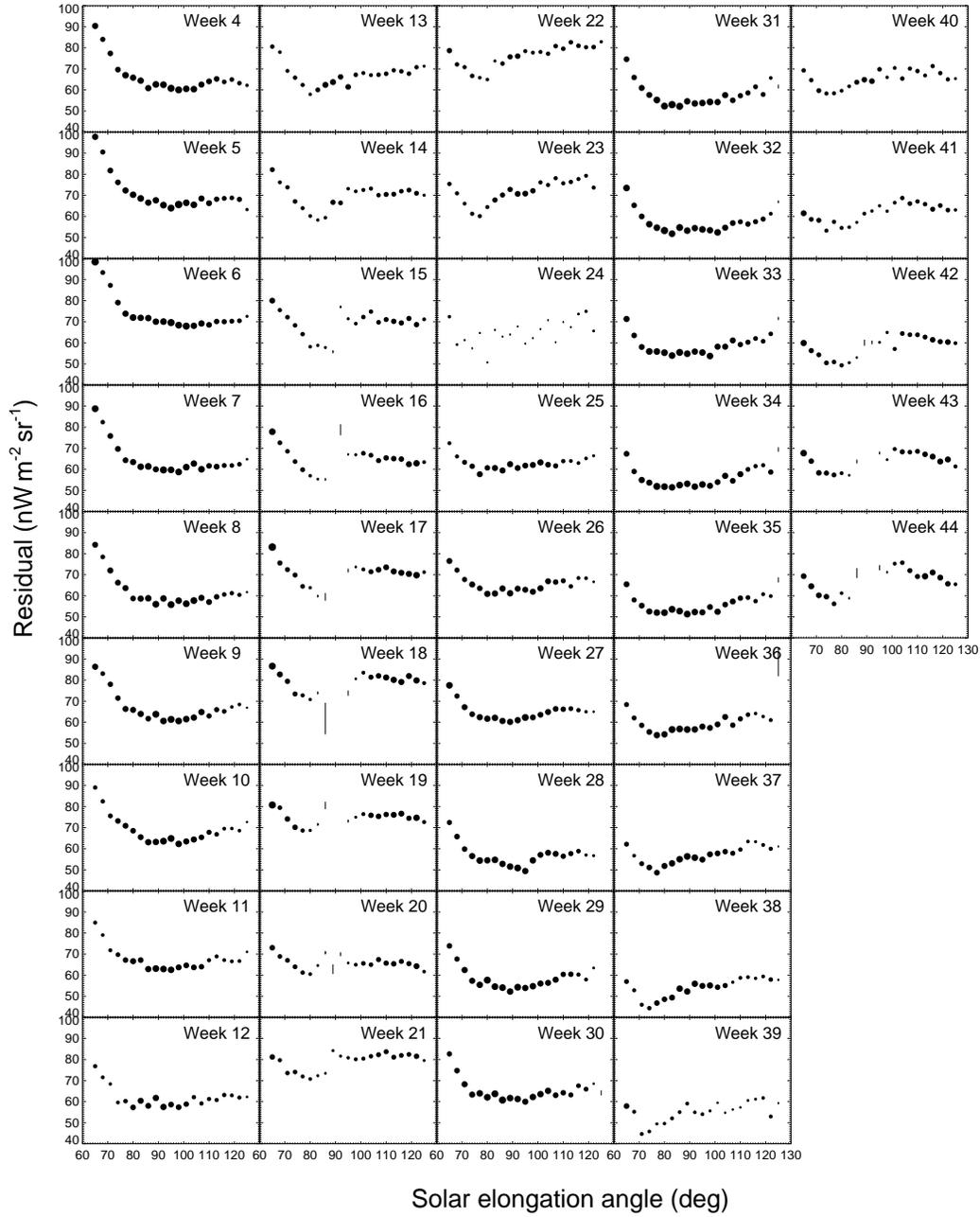}
 \vspace{10mm}
 \caption
 {Solar-elongation dependence of the residual light intensity inferred from each DIRBE weekly map from Week 4 to 44 at $1.25\,{\rm \mu m}$. 
 Black dots represent averaged values of points within each $\Delta\epsilon = 3^\circ$ bin.
Sizes of the dots are set to be proportional to the number of points used to calculate them.
 In some panels, gaps around $\epsilon = 90^\circ$ means no available data points there.
}
\end{center}
\end{figure*}

\begin{figure*}
\begin{center}
 \includegraphics[scale=0.7]{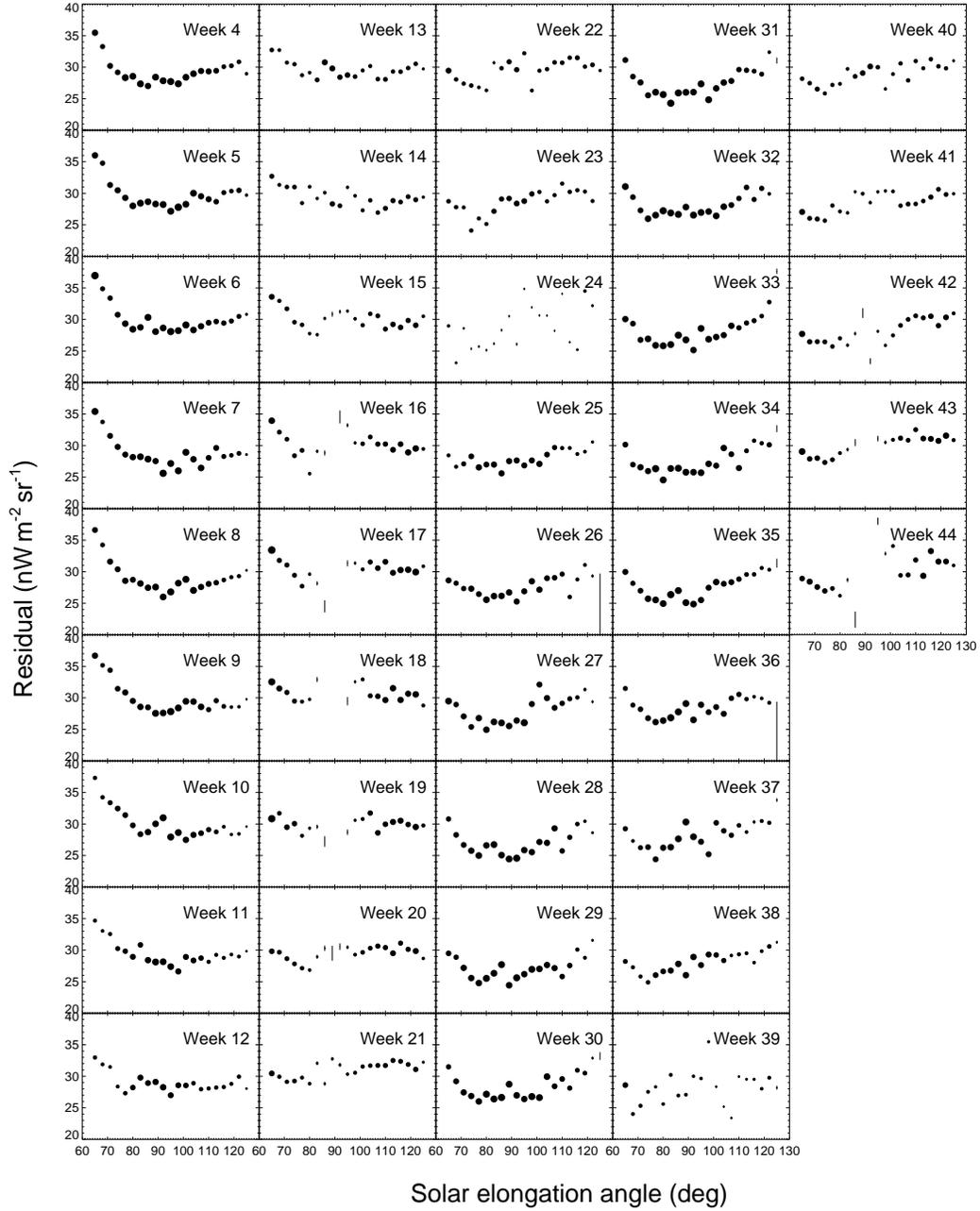}
 \vspace{10mm}
 \caption
 {Same as Fig. 7, but at $2.2\,{\rm \mu m}$. 
}
\end{center}
\end{figure*}

\begin{figure*}
\begin{center}
 \includegraphics[scale=0.7]{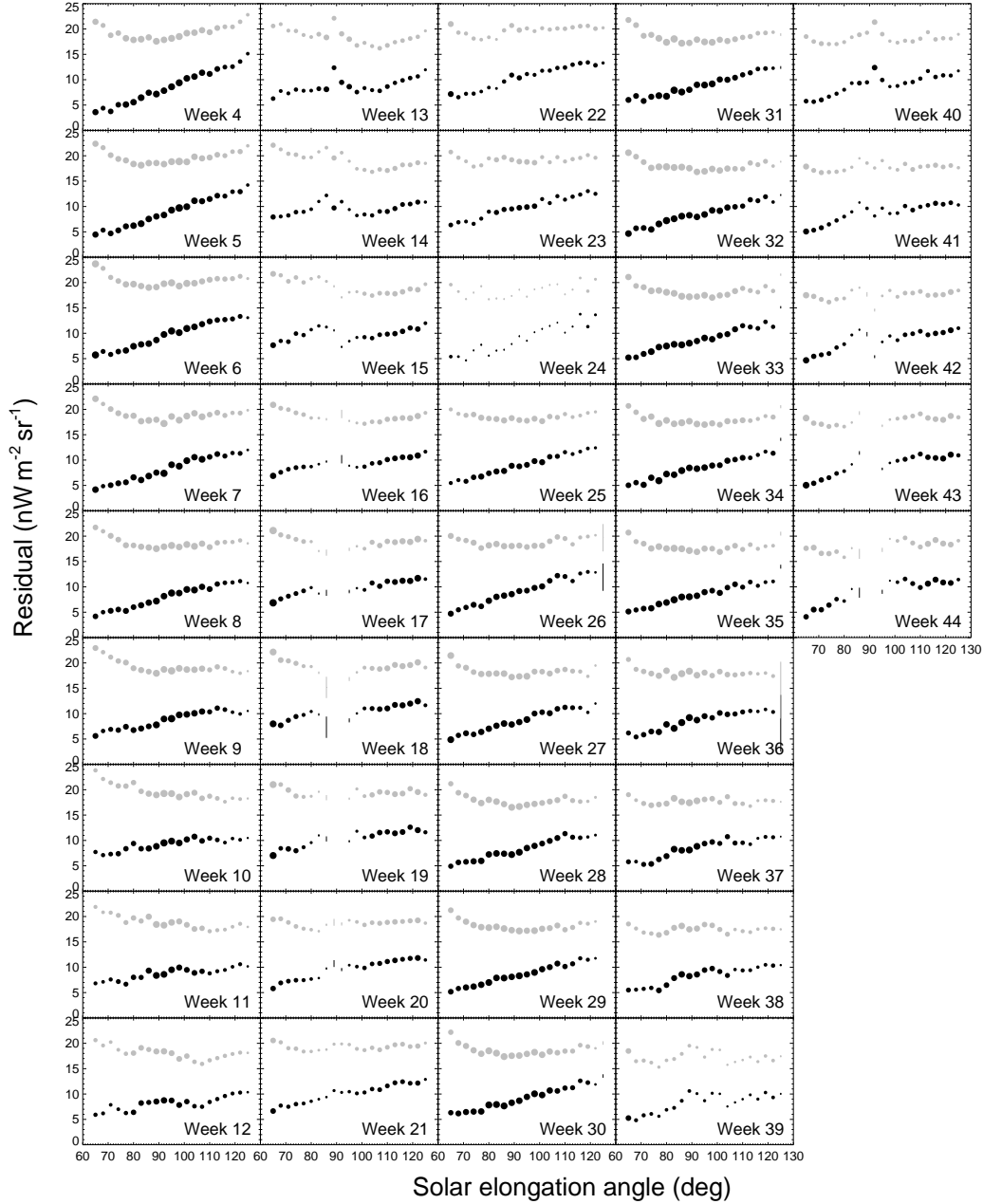}
 \vspace{10mm}
 \caption
 {
 Same as Fig. 7, but at $3.5\,{\rm \mu m}$.
 Black dots represent $\lambda I_{\lambda} ({\epsilon})$ calculated by assuming the $a_\lambda$ value in Table 2, while gray dots indicate the results from the $a_\lambda=1.0$ case (see Section 4.1 and 4.2).
}
\end{center}
\end{figure*}

\begin{figure*}
\begin{center}
 \includegraphics[scale=0.7]{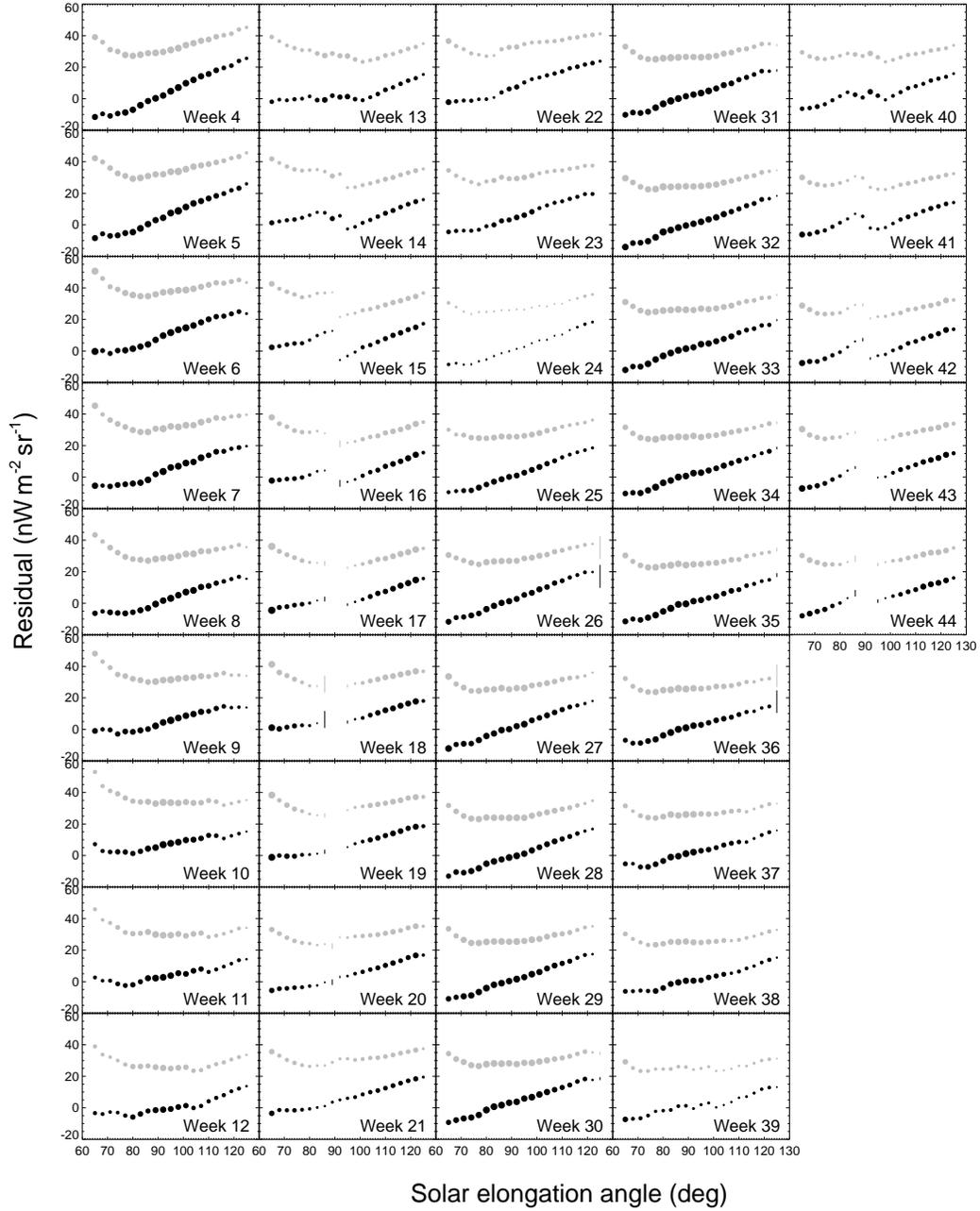}
 \vspace{10mm}
 \caption
 {Same as Fig. 9, but at $4.9\,{\rm \mu m}$. 
}
\end{center}
\end{figure*}

\begin{figure*}
\begin{center}
 \includegraphics[scale=0.7]{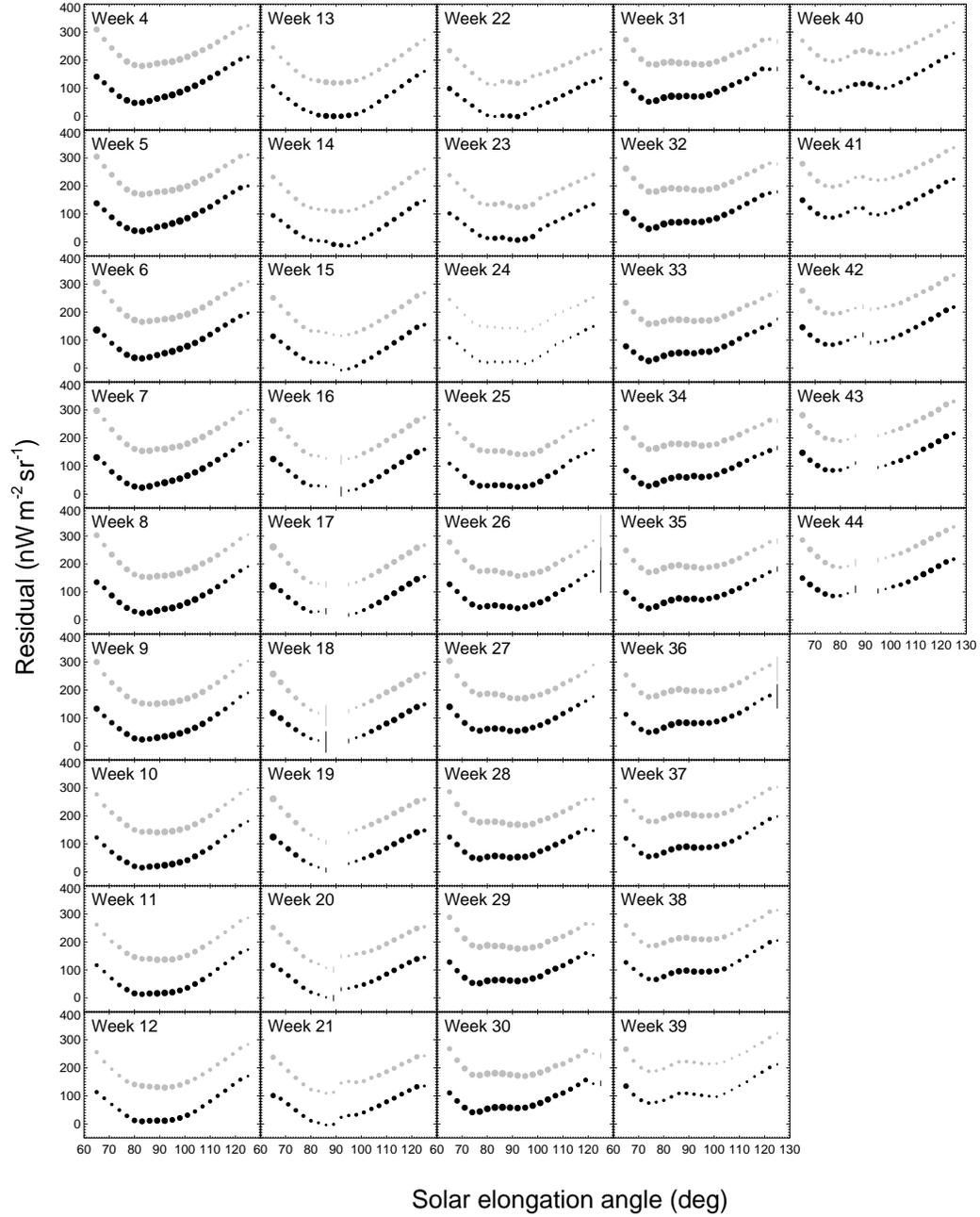}
 \vspace{10mm}
 \caption
 {Same as Fig. 9, but at $12\,{\rm \mu m}$. 
}
\end{center}
\end{figure*}

\begin{figure*}
\begin{center}
 \includegraphics[scale=0.7]{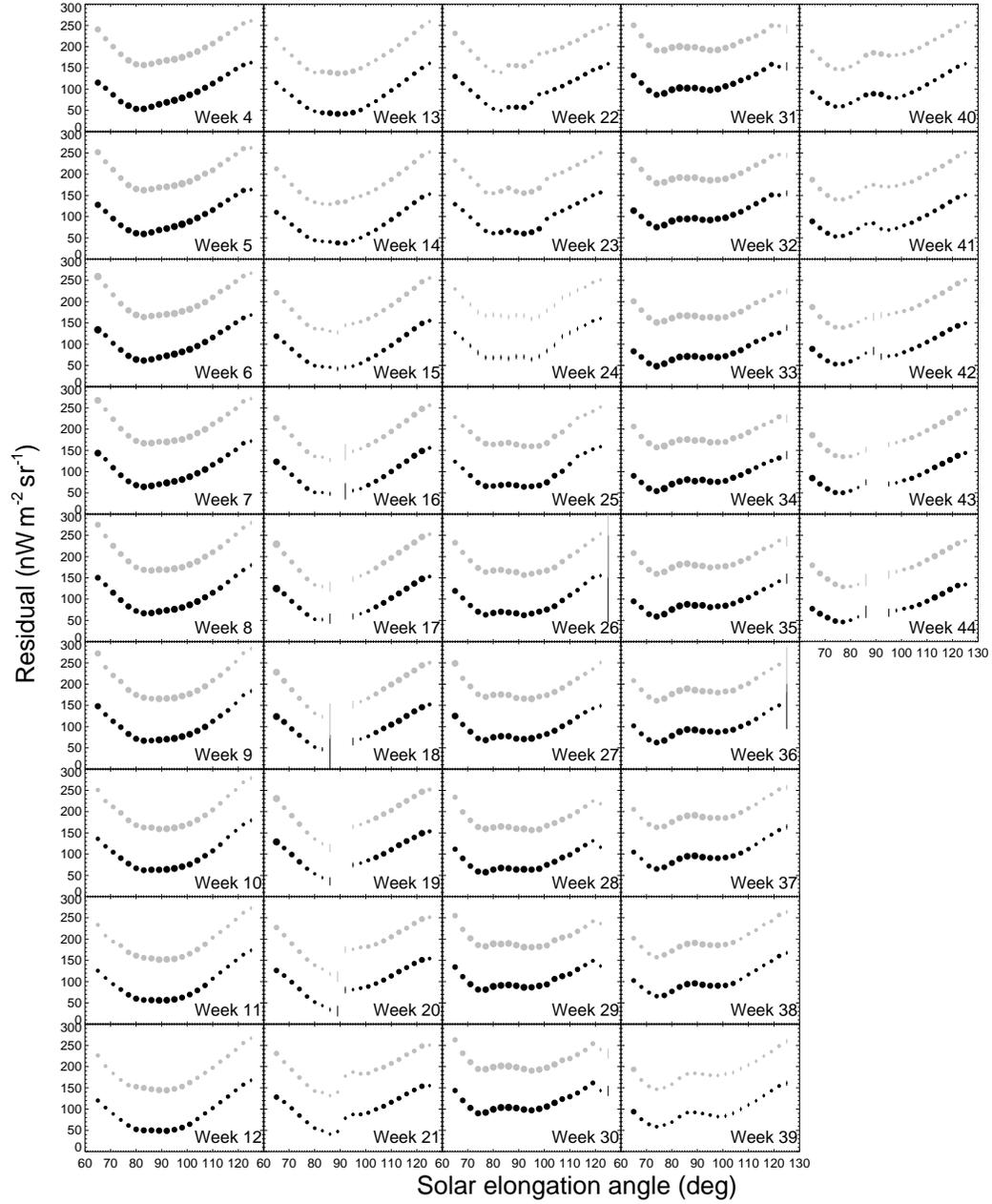}
 \vspace{10mm}
 \caption
 {Same as Fig. 9, but at $25\,{\rm \mu m}$. 
}
\end{center}
\end{figure*}

\begin{figure*}
\begin{center}
 \includegraphics[scale=0.7]{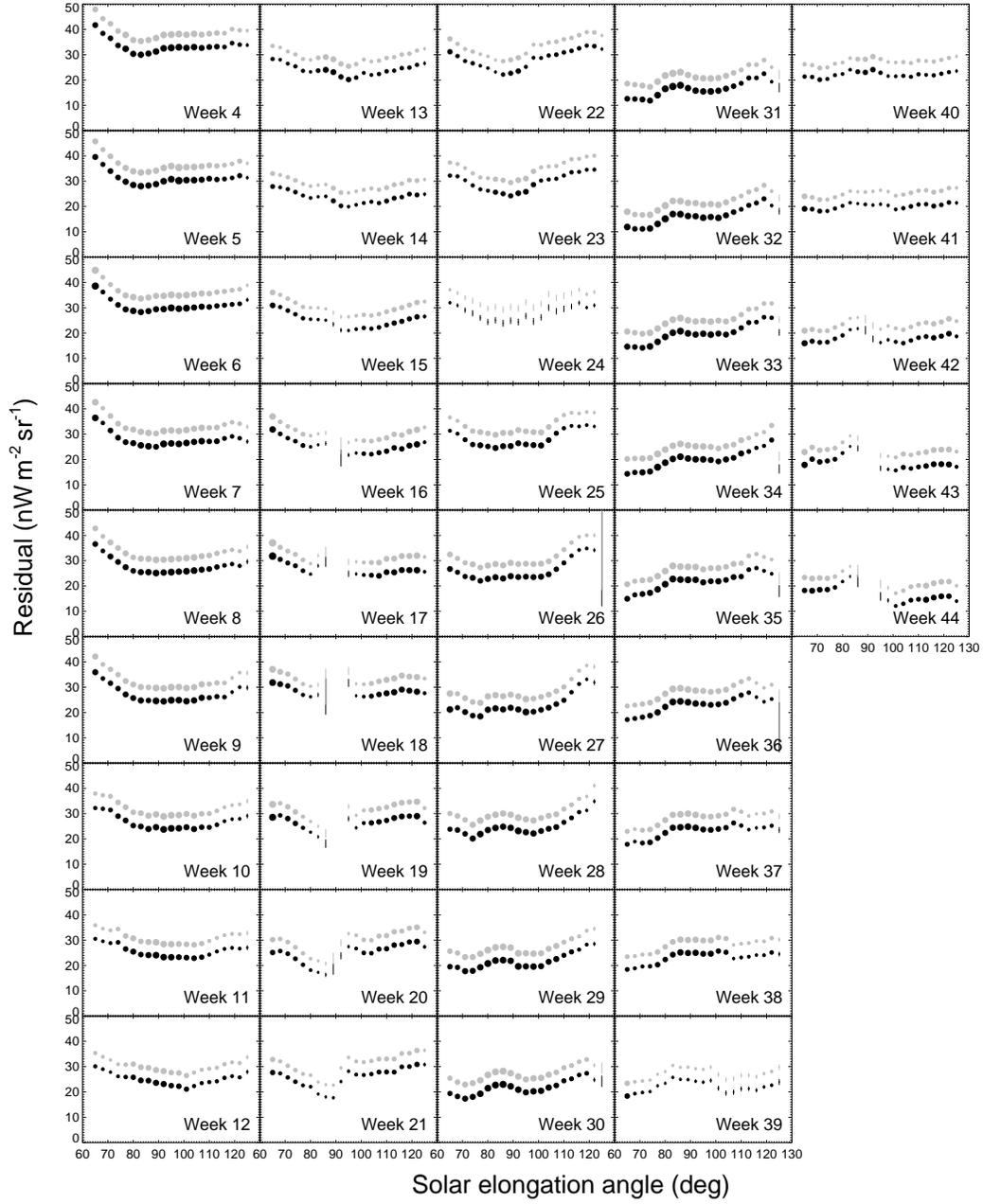}
 \vspace{10mm}
 \caption
 {Same as Fig. 9, but at $60\,{\rm \mu m}$. 
}
\end{center}
\end{figure*}

\clearpage

 \subsection{Comparison with the isotropic IPD model}

As described in Section 4.3, the residual light intensity $\lambda I_{\lambda} (\epsilon)$ shows weekly intensity variation in each band.
Though investigation of the weekly variability of the $\epsilon$-dependence may be important for detailed study of the ZL model, it is hard to trace the features simultaneously in the 7 bands due to the complexity.
Therefore, we conduct comprehensive study of the $\epsilon$-dependence by averaging the results of all the weeks.
Figure 14 shows average weekly averaged $\epsilon$-dependence of Model A or B except for Week 24.
Uncertainty of each point is calculated as quadrature sum of statistical uncertainty of the averaged values and the errors in absolute calibration of the DIRBE observation (Hauser et al. 1998).
On the whole, they show the same $\epsilon$-dependence as those seen in the individual weeks (Fig. 7--12).
At $60\,{\rm \mu m}$, the averaged value is nearly constant because of the large weekly variation in the $\epsilon$-dependence (Fig. 13).
 
 \begin{figure*}
\begin{center}
 \includegraphics[scale=0.8]{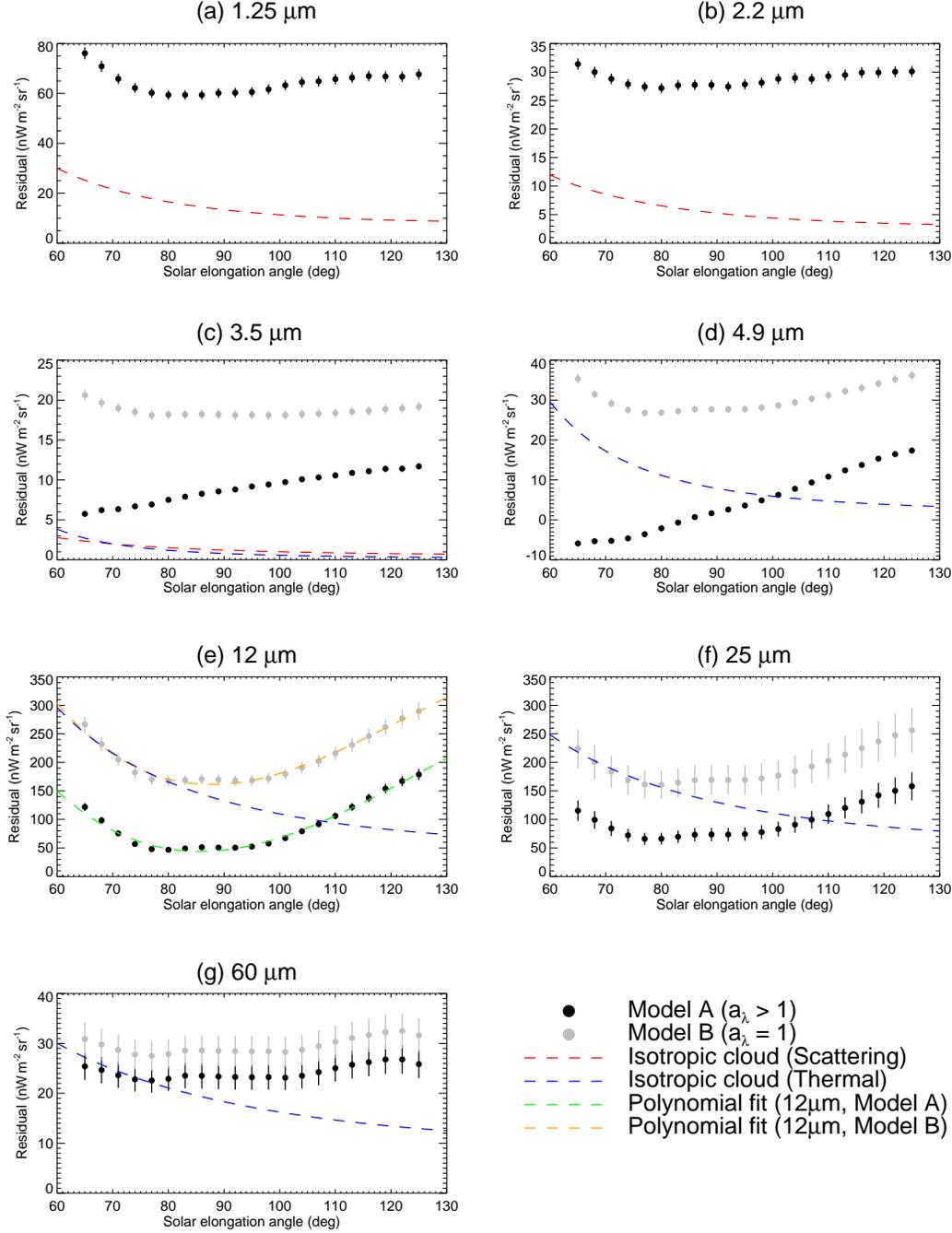}
 \caption
 {Average weekly averaged $\epsilon$-dependence of the residual intensity at (a) $1.25$, (b) $2.2$, (c) $3.5$, (d) $4.9$, (e) $12$, (f) $25$, and (g) $60\,{\rm \mu m}$. 
Results of Model A and B are indicated by black and gray circles, respectively.
Red and blue curves denote, respectively, the $\epsilon$-dependence of the scattered light and thermal emission expected from the isotropic IPD model (Section 2.2) with the number density of $n_0 = 2\times10^{-9}\,{\rm AU^{-1}}$ at $1\,{\rm AU}$ to be fitted to the Model B results in $\epsilon \lesssim 90^\circ$ at $12$ and $25\,{\rm \mu m}$.
Green and orange dashed curves in Panel (e) indicate a polynomial functions fitted to the results of Model A and B, respectively (see Section 6.1 and Table 3).
}
\end{center}
\end{figure*}

To compare with the averaged $\epsilon$-dependence of the residuals, the isotropic IPD models assumed in Section 2.2 are also shown in Fig. 14.
Spatial density distribution of the isotropic IPD is the same as that assumed in Section 2.2 and the other IPD parameters at each wavelength are taken from the Kelsall model (Table 2 of Kelsall et al. 1998).
Density of the isotropic IPD at $1\,{\rm AU}$ is set to fit the residual intensity of Model B in the low-$\epsilon$ regions at $12$ and $25\,{\rm \mu m}$ (Fig. 14).
The density is assumed as $n_0 = 2\times10^{-9}\,{\rm AU^{-1}}$, corresponding to $\sim2\%$ of that of the smooth cloud in the Kelsall model (Table 1). 
From $1.25$ to $25\,{\rm \mu m}$, the $\epsilon$-dependence observed in the low-$\epsilon$ regions is consistent with the isotropic IPD model, except for Model A at $3.5$ to $4.9\,{\rm \mu m}$.
However, difference between the observation and model tends to be large toward the high-$\epsilon$ regions of $\epsilon \gtrsim 90^\circ$.
Though the difference implies that the observed $\epsilon$-dependence cannot be explained by the simple isotropic IPD model, the result indicates presence of an additional component for the IPD.
For simplicity, we continue to call the component  ``isotropic IPD'' in the present paper.

Particularly at $1.25$ and $2.2\,{\rm \mu m}$, residual intensity is a few times higher than that of the isotropic IPD model fitted to the mid-IR residuals (Fig. 14). 
These differences can be regarded as the near-IR EBL component, as discussed in Section 6.

\section{DIFFERENCE BETWEEN THE OBSERVED RESIDUALS AND ISOTROPIC IPD MODEL}

The observed $\epsilon$-dependence of the residual intensity may suggest the presence of the additional IPD component that cannot be described by the simple isotropic IPD model (Section 2.2).
On the other hand, it is possible that some parameters in the Kelsall model are not determined properly since the model includes about fifty physical parameters to represent the ZL brightness.
The brightest component in the Kelsall model, the smooth cloud, can cause the difference between the $\epsilon$-dependence of the residuals and the isotropic IPD model (Fig. 14).
According to the physical representation of the ZL intensity (Equation 1), the spatial density distribution, phase function, and grain temperature of the Kelsall model or the assumed isotropic IPD model may influence the $\epsilon$-dependence. 

\subsection{Spatial distribution of the IPD density}

\begin{figure*}
\begin{center}
 \includegraphics[scale=0.8]{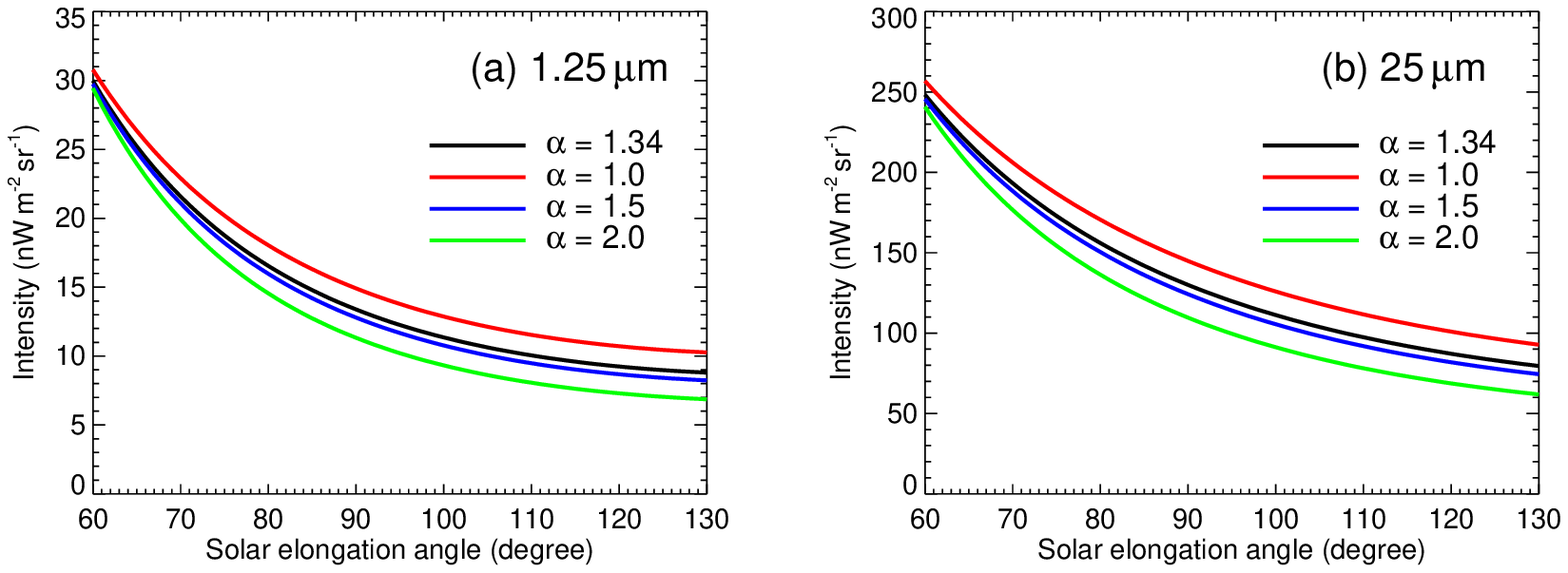}
 \caption
 {Solar elongation dependence of the intensity from the isotropic IPD models with different values of $\alpha$ at (a) $1.25$ and (b) $25\,{\rm \mu m}$ (Section 2.2).
 Black, red, blue, and green curves indicate the models of $\alpha=1.34$, $1.0$, $1.5$, and $2.0$, respectively.
 }
\end{center}
\end{figure*}

We assume the density distribution of the isotropic IPD to have the radial power-law exponent $\alpha=1.34$, same as that of the Kelsall model (Section 2.2).
Since the density distribution of the isotropic IPD is different from the smooth cloud component (Fig. 3), the $\alpha$ value can be different as well.
Earlier observations of the ZL report the $\alpha$ value of $1.0$ to $1.5$ (e.g., Dumont \& S\'anchez 1975; Leinert et al. 1981).
According to the dynamical simulation of Poppe (2016), the dominant ${\rm \mu  m}$-sized IPD density from OCC decreases toward outer solar system with $\alpha \sim 1.0$, which is more gently than that from JFC.

To see influence of the $\alpha$ value on the $\epsilon$-dependence of the ZL intensity from the isotropic IPD, Fig. 15 illustrates the $\epsilon$-dependence at $1.25$ and $25\,{\rm \mu m}$ with four different $\alpha$ values of 1.34, 1.0, 1.5, and 2.0.
These values are motivated by previous studies modeling the IPD (Murdock \& Price 1985; Deul \&  Wolstencroft 1988; Rowan-Robinson et al. 1990; Wheelock et al. 1994; Jones \&  Rowan-Robinson 1993; Kondo et al. 2016).
The isotropic IPD density is assumed as $n_0 = 2\times10^{-9}\,{\rm AU^{-1}}$ to be close to the observed residual level in the mid-IR (Fig. 14) . 
In both wavelengths, the models with the smaller $\alpha$-value show slightly flatter shape toward high-$\epsilon$ regions, but they cannot reproduce the observed $\epsilon$-dependence of the residual intensity (Fig. 14).  
We also confirm that the integration range toward line of sight is not sensitive to the resultant shape of the $\epsilon$-dependence, though we do not show it explicitly.

If the density of the isotropic IPD increases as a function of $R$, it may reproduce the observed $\epsilon$-dependence of the residual intensity.  
Considering both Poynting-Robertson drag and solar radiation pressure, Poppe (2016) presents dynamical simulation of the IPD  grains, assuming grain size distribution of $dn/da \propto a^{-2.5}$ with the radii $a$. 
According to their result, the density of the OCC grains with $a \gtrsim 20\,{\rm \mu m}$ increases toward regions of $R\sim20\,{\rm AU}$, while that of the ${\rm \mu m}$-sized grains shows the $R^{-1}$ dependence (Fig. 6 of Poppe 2016). 
The $R$-dependence of the larger grains can be caused by increasing contribution of the solar radiation pressure to the large dust.
However, the density of the large grains is expected to be less than $1\%$ of that of the ${\rm \mu m}$-sized grains (Fig. 6 of Poppe 2016).
This indicates negligible contribution of the large grains to the $\epsilon$-dependence. 
According to these discussions, the $\alpha$ value of the isotropic IPD model seems difficult to explain the observed $\epsilon$-dependence
of the residual intensity, particularly the features seen in the high-$\epsilon$ regions at $12$ and $25\,{\rm \mu m}$.

Uncertainty of the spatial density distribution in the Kelsall model, particularly that of the smooth cloud component, could affect the observed $\epsilon$-dependence of the residuals as well.
The Kelsall model comprises too many geometric parameters to investigate the sensitivity of each parameter to the $\epsilon$-dependence (e.g., Equation 2).
However, the geometric parameters are determined to fit the DIRBE maps of $10$ photometric bands, indicating more reliability than the parameters related to either scattered light or thermal emission, such as phase function or grain temperature.  
Therefore, we do not investigate the geometric parameters of the Kelsall model in the present paper.

\subsection{Scattering phase function and albedo}

For the scattered light, the phase function can affect the $\epsilon$-dependence, as inferred from Equation (1).
Figure 16(a) shows the phase function derived in the Kelsall model at $1.25$, $2.2$, and  $3.5\,{\rm \mu m}$ (Equation 3), which is also adopted to calculate the scattered light from the isotropic IPD (Section 2.2).
From Equation (5) and (6), Fig. 16(b) illustrates the scattering angle ($\theta$) as a function of solar elongation angle ($\epsilon$) an IPD grain at the position $s$ (Fig. 2).
According to Fig. 2b, the $\epsilon$ range $64^\circ \lesssim \epsilon \lesssim 124^\circ$ of the DIRBE map corresponds to $60^\circ \lesssim \theta \lesssim 150^\circ$ for the IPD at the $s \lesssim 1.0\,{\rm AU}$ regions where the ZL contribution is dominant.
Therefore, shape of the phase function in the $\theta$ range is expected to influence the $\epsilon$-dependence of the scattered light.
In the $\theta$ range, the phase function is similar among the three wavelengths.
Since the $\epsilon$-dependence of the residual intensity at $1.25$, $2.2$, and  $3.5\,{\rm \mu m}$ in Model B is also similar to each other (Fig. 14), the phase function may partly cause the difference between the observation and the isotropic IPD model in the high-$\epsilon$ regions.


Some studies report $R$-dependence of the grain albedo (Lumme \& Bowell 1985; Renard et al. 1995).
This indicates that size distribution or composition of the IPD also changes as a function of $R$.
Though these IPD properties are expected to influence the $\epsilon$-dependence of the ZL, we do not discuss it here due to the complication.

\begin{figure*}
\begin{center}
 \includegraphics[scale=0.7]{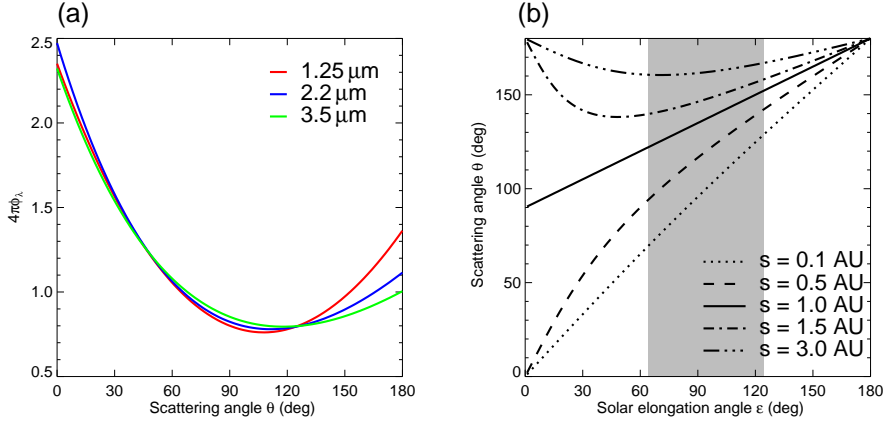}
 \caption
 {Panel (a): Parameterized scattering phase function derived in the Kelsall model (Equation 3).
Red, blue, and green curves indicate the phase function at $1.25$, $2.2$, and $3.5\,{\rm \mu m}$, respectively. 
Panel (b): Relation between solar elongation angle ($\epsilon$) and scattering angle ($\theta$) for a grain located at $s$, inferred from Equation (5) and (6).
Dotted, Dashed, solid, dot-dashed, and dot-dot-dot-dashed curves indicate the cases of $s=0.1$, $0.5$, $1.0$, $1.5$, and $3.0\,{\rm AU}$, respectively.
A shaded region represents the $\epsilon$ range of the DIRBE observation ($64^\circ \lesssim \epsilon \lesssim 124^\circ$).
}
\end{center}
\end{figure*}

\subsection{The IPD temperature as a function of heliocentric distance}

The $R$-dependence of the grain temperature is expected to affect the $\epsilon$-dependence of the thermal emission.
In the Kelsall model, the temperature power-law exponent $\delta$ is assumed as $\delta=0.467$.
In general, equation of the thermal equilibrium for a dust grain of size $a$ at heliocentric distance $R$ is given by
$$ 
\int  \pi a^2 Q_{\lambda, {\rm abs}} (a) \pi (R_\odot/R)^2 B_{\lambda} (T_\odot) d\lambda
$$
\begin{equation}
 = \int  \pi a^2 Q_{\lambda, {\rm abs}} (a) 4\pi  B_{\lambda} [T_g(R)] d\lambda,
\end{equation}
where $Q_{\lambda, {\rm abs}} (a)$, $R_\odot$ , $T_\odot$, and $T_g(R)$ are absorption coefficient of the grain, solar radius, solar temperature, and grain temperature at $R$, respectively.
To test validity of $\delta=0.467$ derived in the Keslall model, we calculate $T_g(R)$ in the case of $a=0.1$, $1.0$, and $10\,{\rm \mu m}$ from Equation (14), assuming $Q_{\lambda, {\rm abs}} (a)$ of spherical silicate grains (Draine \& Lee 1984 and Laor \&  Draine 1993).

Figure 17(a) shows the results of $T_g(R)$ for the different grain sizes.
For comparison, the analytical forms $T = T_0 R^{-\delta}$ (Equation 4) with $\delta=0.467$, $0.4$, $0.5$, and $0.6$ are plotted as well.
The curve with $\delta=0.467$ runs between the models of $a=1.0$ and $10\,{\rm \mu m}$.
Though composition or shape of the IPD grains should be taken into account for more detailed discussion, this test indicates that the averaged IPD size is approximately a few micrometer.
This is marginally consistent with earlier studies of the IPD (e.g., Poppe 2016; Fixsen \& Dwek 2002).

To see sensitivity of $\delta$ for the ZL intensity, Fig. 17(b) describes $\epsilon$-dependence of the thermal emission intensity from the smooth cloud in the Kelsall model at $25\,{\rm \mu m}$ with the $\delta$ values of $0.467$, $0.4$, $0.5$, and $0.6$. 
Since the smooth cloud component shows the elongated structure toward the ecliptic plane (Fig. 3a), total dust density toward line of sight tends to increase as a function of $\epsilon$ in high-$\epsilon$ regions.
This effect causes the intensity rise in the regions of $\epsilon \gtrsim 110^\circ$ in any $\delta$ (Fig. 17b).
Intensity difference among these models is smaller in the low-$\epsilon$ regions, but it is larger toward the high-$\epsilon$ regions ($\epsilon \gtrsim 90^\circ$).
If we adopt the model with $\delta=0.4$ instead of the default value of $0.467$, the ZL intensity is about $100\,{\rm nW\,m^{-2}\,sr^{-1}}$ higher in the high-$\epsilon$ regions.
This amount is comparable to the difference between the observed residuals and the isotropic IPD model at $25\,{\rm \mu m}$ (Fig. 14f). 
Therefore, the $R$-dependence of the grain temperature with $\delta=0.4$ can partly explain the inconsistency in the mid-IR high-$\epsilon$ regions.     
According to analysis of the {\it IRAS} data, Wheelock et al. (1994) derive the small value of $\delta=0.36$, closer to $0.4$. 
Since the low value of $\delta$ implies more amount of sub-micrometer- or micrometer-sized grains (Fig. 17a), this may indicate that the density of the smaller IPD grains is higher than that assumed in the Kelsall model.

\begin{figure*}
\begin{center}
 \includegraphics[scale=0.8]{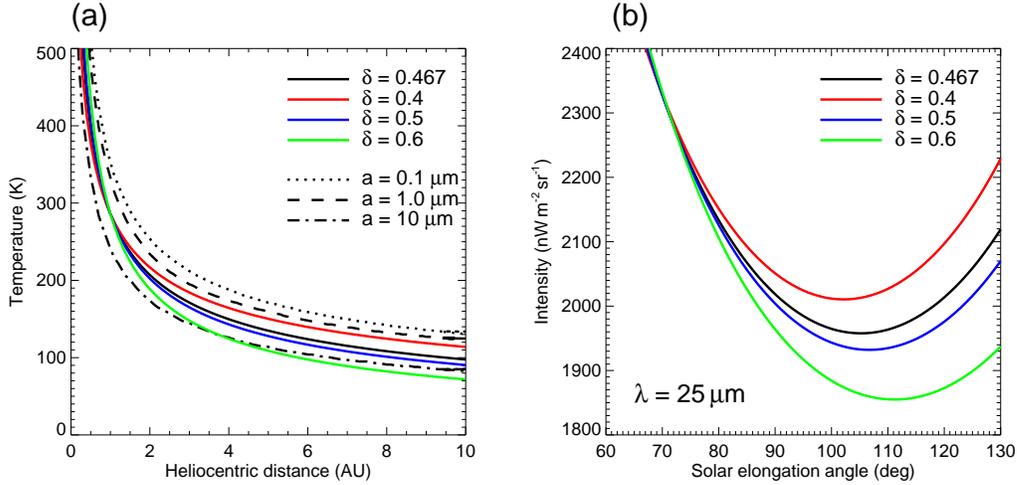}
 \caption
 {Panel (a): Grain temperature as a function of heliocentric distance $R$.
 Black dotted, dashed, and dot-dashed curves represent the results for silicate grains of the size $a=0.1$, $1.0$, and $10\,{\rm \mu m}$, respectively.
 Black, red, blue, and green solid lines represent, respectively, an analytic form of $T=T_0 R^{-\delta}$ with $\delta = 0.467$ (Kelsall et al. 1998), $0.4$, $0.5$, and $0.6$.
 Panel (b): Solar elongation-dependence of the ZL intensity of the smooth cloud in the Kelsall model at $25\,{\rm \mu m}$ with $\delta = 0.467$ (black), $0.4$ (red), $0.5$ (blue), and $0.6$ (green).
}
\end{center}
\end{figure*}
  
\section{SEPARATION OF THE EBL FROM THE ISOTROPIC IPD COMPONENT}

The residual intensity $\lambda I_{\lambda} (\epsilon)$ derived in Section 4 should include the isotropic IPD component and the EBL independent of $\epsilon$.
In the mid-IR, the IGL intensity derived from deep galaxy counts is an order of a few ${\rm nW\,m^{-2}\,sr^{-1}}$ (Elbaz et al. 2002; Papovich et al. 2004), which is  lower than the residual intensity by two orders of magnitude (Fig. 1 and Fig. 14).
Due to no observational evidence of high intensity of the EBL in the mid-IR, it is reasonable to assume the EBL to be the same level as the IGL.
Then, the entire residual intensity is thought to originate from the ZL component in the mid-IR. 

\subsection{Spectral energy distribution of the residuals and density of the isotropic IPD component}

\begin{figure*}
\begin{center}
 \includegraphics[scale=0.8]{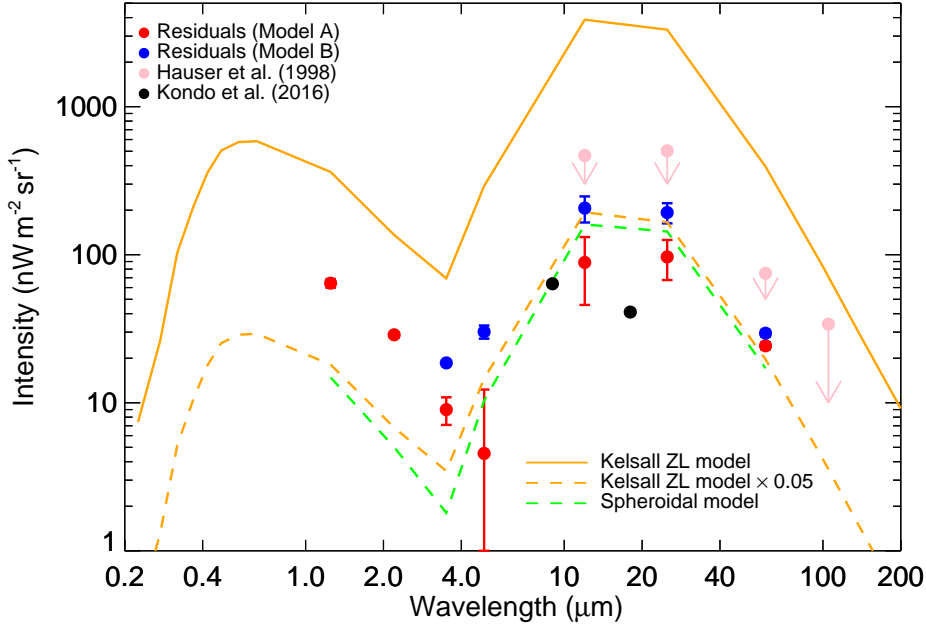}
 \caption
 {Spectral energy distribution of the averaged residual intensity derived in the present study.
 Results from Model A and B are indicated by red and blue circles, respectively.
 Upper limits of the  EBL are represented by pink allows (Hauser et al. 1998).
 Black circles at $9$ and $18\,{\rm \mu m}$ denote the isotropic component derived from the ZL model of the {\it AKARI} all-sky map (Kondo et al. 2016).
 A solid orange curve indicates the ZL spectrum from the Kelsall model and Kawara et al. (2017) in the intermediate ecliptic latitudes, same as Fig. 1, while dashed one denotes $5\%$ level of that to be comparable to the residual intensity at $12$ and $25\,{\rm \mu m}$.
A green dashed curve denotes the SED expected from the spheroidal IPD model fitted to the observed $\epsilon$-dependence of the residual intensity at $12\,{\rm \mu m}$ (Section 6.2).   
 }
\end{center}
\end{figure*}

Figure 18 shows spectral energy distribution (SED) of the derived residuals from near- to mid-IR.
The residual values are calculated as the average of the $\epsilon$-dependence at each wavelength (Fig. 14).
The orange dashed line is a scaled spectrum of the Kelsall ZL model, fitted to the mid-IR residuals.
At $12$, $25$, and $60\,{\rm \mu m}$, color of the residuals is marginally consistent with that of the Kelsall model.
This may indicate that the entire residuals originate from the isotropic IPD component missed in the Kelsall model.
At $1.25$ and $2.2\,{\rm \mu m}$, the residuals are several times larger than the scaled ZL spectrum.
The difference between them can be regarded as the EBL in the near-IR.   

We can estimate the density of the isotropic IPD from the mid-IR residual intensity.
Regardless of the difference between Model A and B at $12$ and $25\,{\rm \mu m}$ (Fig. 18), they are close to the $5\%$ ZL intensity of the Kelsall model.
This indicates that the density of the isotropic IPD is also an order of $5\%$ of that of the dominant smooth cloud in the Kelsall model (Table 1).
As shown in Section 4.4, the $\epsilon$-dependence of the mid-IR residuals in the low-$\epsilon$ regions is close to the simple isotropic IPD model whose density is $\sim2\%$ of the Kelsall model.
However, this estimate underestimates the density of the isotropic IPD because of the different trends between the observation and model in the high-$\epsilon$ regions (Fig. 14).
Therefore, we assume the density fraction of the isotropic IPD to be $\sim5\%$ of the total IPD.
This value is consistent with other studies suggesting the mass fraction of the OCC dust is less than $\sim10\%$ of that of the total IPD (Hahn et al. 2002; Nesvorn\'y et al. 2010; Poppe 2016).

\subsection{Contribution of the isotropic IPD at the near-IR wavelengths}

To quantify the contribution of the ZL intensity of the isotropic IPD in the near-IR from the mid-IR result, we fit the $\epsilon$-dependence of the residuals at $12\,{\rm \mu m}$ by a cubic polynomial function, 
\begin{equation}
 f(\epsilon) = a_0 + a_1 \epsilon + a_2 \epsilon^2 + a_3 \epsilon^3,
 \end{equation}
where $a_0$, $a_1$, $a_2$, and  $a_3$ are free parameters determined by the fitting.
The EBL (IGL) contribution is assumed to be negligible against the residual intensity at $12\,{\rm \mu m}$.
Therefore, the polynomial fitting at $12\,{\rm \mu m}$ is conducted without subtracting the IGL component from the residual intensity.
The fitting results for Model A and B at $12\,{\rm \mu m}$ are shown in Table 3 and the polynomial functions are plotted in Fig. 14 with the residuals at $12\,{\rm \mu m}$.

\begin{table*}
\begin{center}
 \renewcommand{\arraystretch}{1.0}
 \caption{Results of cubic polynomial fit to the $\epsilon$-dependence of the residuals at $12\,{\rm \mu m}$ (Equation 15)}
  \label{symbols}
  \begin{tabular}{lcccc}
  \hline
   Parameters & $a_0\,({\rm nW\,m^{-2}\,sr^{-1}})$ & $a_1\,({\rm nW\,m^{-2}\,sr^{-1}\,deg^{-1}})$ & $a_2\,({\rm nW\,m^{-2}\,sr^{-1}\,deg^{-2}})$ & $a_3\,({\rm nW\,m^{-2}\,sr^{-1}\,deg^{-3}})$ \\
 \hline
   Model A ($a_\lambda > 1$)      & $1712$ & $-47.06$ & $0.4169$ & $-0.001108$ \\
   Model B ($a_\lambda=1$)     & $2242$ & $-58.43$ & $0.5208$ & $-0.001426$ \\
          \hline
        \end{tabular}
    \end{center}
    \medskip
 \end{table*}
 
 \begin{figure*}
\begin{center}
 \includegraphics[scale=0.8]{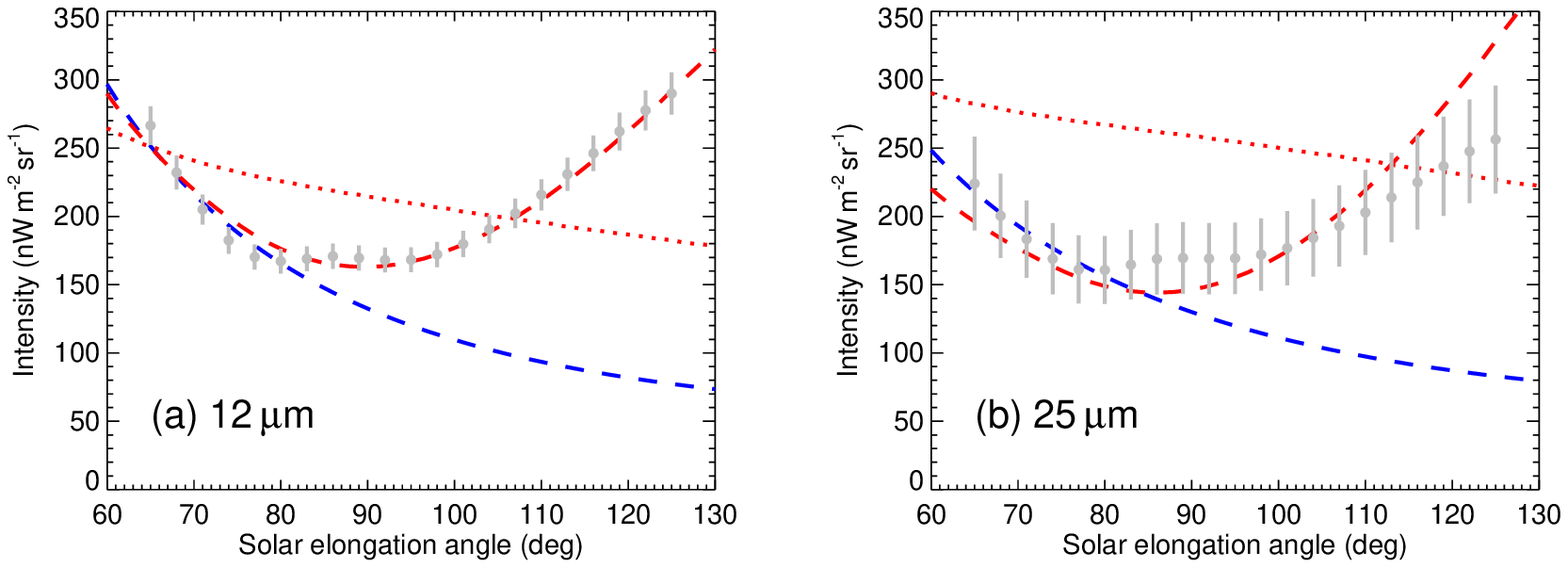}
 \caption
 {Comparison of models of the isotropic IPD cloud and $\epsilon$-dependence of the residuals obtained at $12\,{\rm \mu m}$ (Panel a) and $25\,{\rm \mu m}$ (Panel b).
 Red dashed curves indicate the spheroidal model ($e=0.9$) of two density components, fitted to the $\epsilon$-dependence of the residuals (see Section 6.2 for the model description).
Red dotted curves represent the spherical model ($e=0$) by assuming the same $R$-dependence of the density as the spheroidal model. 
The intensity of the spherical model is scaled by a factor of $0.3$ to be comparable to the other values.
The $\epsilon$-dependence of the residuals derived from Model B are indicated by gray circles, same as Fig. 14. 
Blue dashed curves denote the $\epsilon$-dependence expected from the simple isotropic IPD model (Section 2.2).
}
\end{center}
\end{figure*}

To explain the observed $\epsilon$-dependence at $12\,{\rm \mu m}$ (Fig. 14), we search for a new model of the isotropic IPD.
 As inferred from Equation (1), the density distribution reproducing the $\epsilon$-dependence should be different from that assumed in the initial prediction of the isotropic IPD (Section 2.2).
 We then assume $R$-dependence of the IPD density as
   \begin{equation}
n(R) \propto \left \{
\begin{array}{ll}
R^{-1} &  \mbox{($R \lesssim R_{\rm th}$)}\\
R^{\gamma}  & \mbox{($R \gtrsim R_{\rm th}$),}
\end{array}
\right.
\end{equation}
where $\gamma$ and $R_{\rm th}$ are parameters and $n(R)$ is continuous at $R_{\rm th}$.
We also allow the cloud shape to be spheroid to describe deviation from sphere.
Parameters of the ellipse from which the spheroid originates are characterized by the major axis $2A$ and eccentricity $e$.
The major axis of the ellipse is assumed to be on the ecliptic plane.
The other IPD parameters except for the density distribution are set as those derived in the Kelsall model.
We search for the spheroidal models fitted to the $\epsilon$-dependence of the residuals at $12\,{\rm \mu m}$ by changing these parameters.

In Figure 19, red dashed curves indicate one spheroidal model fitted to the residuals of Model B at $12$ and $25\,{\rm \mu m}$.
 The model reproduces the $\epsilon$-dependence particularly in high-$\epsilon$ regions, where the spherical cloud assumed in Section 2.2 (blue dashed curves) cannot make the behavior.
 The parameters of the spheroidal model are $\gamma=10$, $R_{\rm th}=1.31\,{\rm AU}$, $A=2\,{\rm AU}$, and $e=0.9$ with the density at $R = 1\,{\rm AU}$ being $2.5\times 10^{-9}\rm AU^{-1}$, which indicates drastic increase of the IPD density in farther regions from the sun.
If the cloud shape is sphere (i.e., $e=0$) with the other parameters same as the spheroidal model, the $\epsilon$-dependence becomes far from the observed residuals (red dotted curves in Fig. 19).
Since the spheroidal model includes several parameters, the assumed parameter values should not be a unique representation of the residuals. 
This exercise indicates that some models can reproduce the observed $\epsilon$-dependence at $12\,{\rm \mu m}$. 
In addition, the uncertainty factors in the Kelsall model may contribute the $\epsilon$-dependence (Section 5).
   
Assuming the SED of the ZL intensity from the isotropic IPD to be same as that of the Kelsall model, we estimate the $\epsilon$-dependence of the intensity in the other wavelengths according to the fitting results at $12\,{\rm \mu m}$. 
The ZL intensity ratios of the other wavelengths to the $12\,{\rm \mu m}$, $C_\lambda$ in units of ${\rm (nW\,m^{-2}\,sr^{-1})/(nW\,m^{-2}\,sr^{-1})}$, can be calculated from the SED of the Kelsall model (Fig. 18) and the $a_\lambda$ values, the scaling factor against the Kelsall model (Table 2).
Table 4 lists the derived values of $C_\lambda$ for Model A and B.
Then, the ZL intensity of the isotropic IPD can be calculated as $C_\lambda a_\lambda f(\epsilon)$. 
The intensity obtained by subtracting the isotropic IPD component from the residuals corresponds to the EBL component.

The SED expected from the spheroidal IPD model (Fig. 19) should be different from that of the Kelsall model due to the difference in the density distribution.
To show the SED difference, we calculate $C_\lambda$ from the spheroidal model and list them in Table 4 as ``Spheroidal model'' by assuming the same IPD properties as the Kelsall model except for the density distribution.
The SED of the spheroidal model is also shown in Fig. 18.
The result shows that the spheroidal model changes the $C_\lambda$ values against Model B by $\sim2\%$ ($1.25\,{\rm \mu m}$) to $\sim40\%$ ($3.5\,{\rm \mu m}$).
The SED difference is included in the uncertainty of the near-IR EBL (Section 6.3).

Figure 20 shows the $\epsilon$-dependence derived by subtracting the isotropic IPD component $C_\lambda a_\lambda f(\epsilon)$ from the residuals (Fig. 14).
The results obtained from Model A and B at $12\,{\rm \mu m}$ are indicated by red and blue dots, respectively.
Nearly zero values at $12\,{\rm \mu m}$ indicate good fitting to the residual intensity by the polynomial functions (Equation 15).
At the other wavelengths, the $\epsilon$-dependence becomes weaker than that of the  residuals (Fig. 14).
This indicates that the $\epsilon$-dependence of the ZL intensity from the isotropic IPD is similar to the polynomial functions derived at $12\,{\rm \mu m}$.
Particularly at $1.25$ and $2.2\,{\rm \mu m}$, the results from both Model A and B still leave high intensity.
These components can be regarded as the EBL in the near-IR, as is also inferred from the intensity difference between the scaled ZL and residuals in Fig. 18.
Results of the $\epsilon$-averaged residuals, ZL intensity from the isotropic IPD, and their difference (EBL) obtained from Model A and B are listed in Table 5.

\begin{table*}
\begin{center}
 \renewcommand{\arraystretch}{1.0}
 \caption{Intensity ratios of the ZL to $12\,{\rm \mu m}$, $C_\lambda$ in units of ${\rm (nW\,m^{-2}\,sr^{-1})/(nW\,m^{-2}\,sr^{-1})}$}
  \label{symbols}
  \begin{tabular}{lccccccc}
  \hline
   Band ($\rm{\mu m}$) & $1.25$ & $2.2$ & $3.5$ & $4.9$ & $12$ & $25$ & $60$ \\
 \hline
      Model A   ($a_\lambda >1$)   & $0.0911$ & $0.0355$ & $0.0199$ & $0.0800$ & $1.0$ & $0.855$ & $0.0998$\\
      Model B   ($a_\lambda=1$)   & $0.0944$  &  $0.0367$ &   $0.0179$ &   $0.0753$   &   $1.0$  &   $0.856$  &   $0.102$ \\
      Spheroidal model   & $0.0926$  &  $0.0313$ &   $0.0113$ &   $0.0646$   &   $1.0$  &   $0.896$  &   $0.107$ \\
          \hline
        \end{tabular}
    \end{center}
    \medskip
 \end{table*}

\begin{figure*}
\begin{center}
 \includegraphics[scale=0.8]{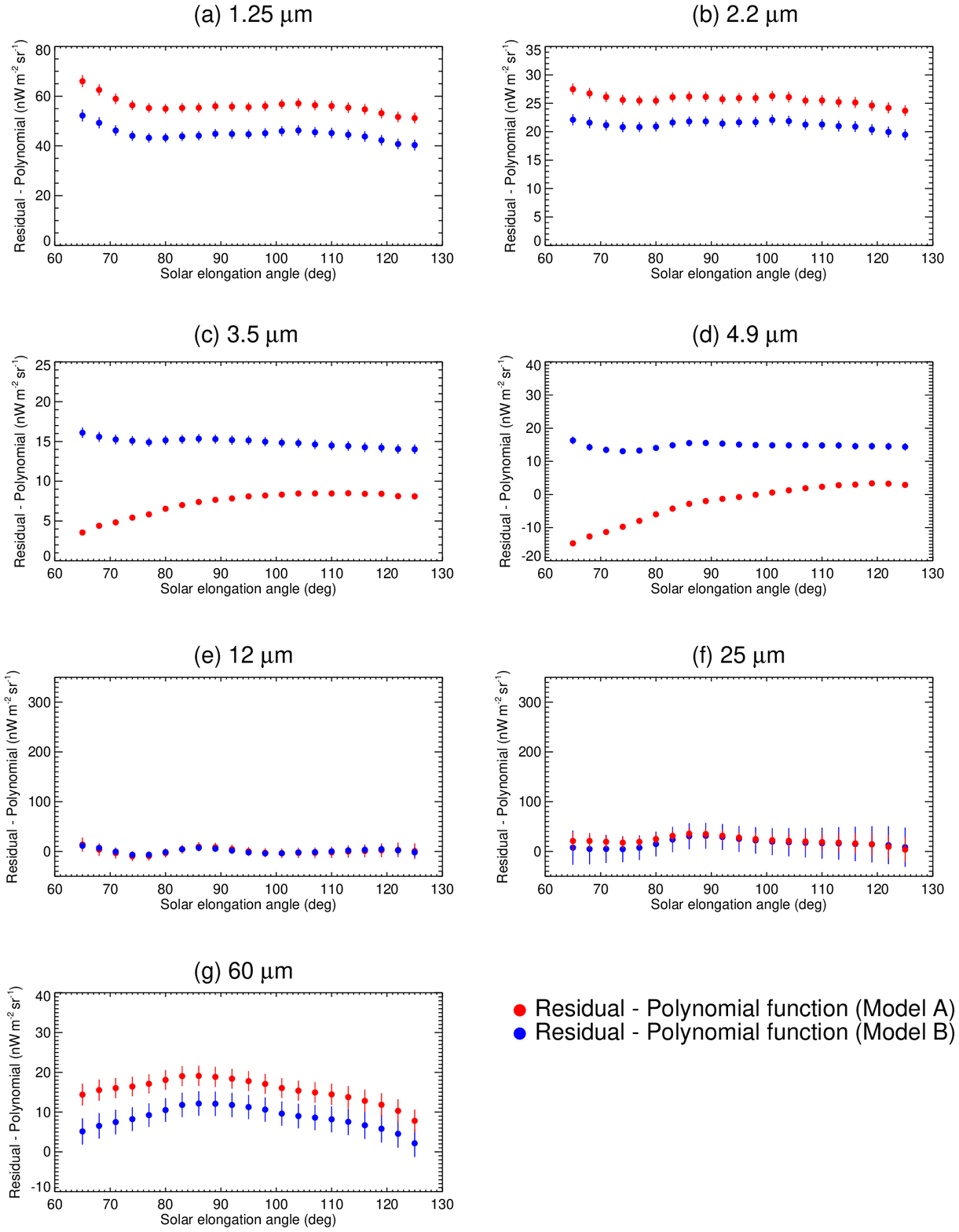}
 \caption
 {Solar elongation dependence of the intensity derived by subtracting the polynomial function fitted to the residual intensity at $12\,{\rm \mu m}$ (Equation 15) from the residuals at (a) $1.25$, (b) $2.2$, (c) $3.5$, (d) $4.9$, (e) $12$, (f) $25$, and (g) $60\,{\rm \mu m}$ (Fig. 14).
Red and blue dots indicate the results in Model A and B, respectively.
}
\end{center}
\end{figure*}

\subsection{Uncertainties of the resultant EBL intensity}

Several uncertainties should be taken into account to obtain the resultant EBL intensity.
The uncertainty components listed in Table 5 are estimated as follows.  

Uncertainties of absolute gain calibration of the DIRBE observations are derived in Hauser et al. (1998).
At each band, the uncertainty is given in units of percentage.
These uncertainties are calculated as the percentage of the residual intensity and listed as `` Absolute calibration'' in Table 5.

A little $\epsilon$-dependence still remains in the EBL intensity after subtracting the isotropic IPD component from the residuals (Fig. 20).
The peak-to-valley values of the $\epsilon$-dependence are included in the uncertainty.
These values are listed as ``$\epsilon$-dependence'' in Table 5.
The ``upper'' and ``lower'' values indicate, respectively, the maximum minus the averaged values and the averaged minus the minimum values of ``Residuals - Isotropic IPD'' (Fig. 20).

Uncertainties from statistics and DGL $b$-dependence are common for the results from Model A and B.
Statistical uncertainties are estimated as the average of those calculated in the individual $\epsilon$ bins of the residuals (Fig. 14).
The values are listed as ``Statistical'' in Table 5.
As noted in Section 4, our model of the sky brightness does not include the $b$-dependence of the DGL (Equation 9).
Sano \& Matsuura (2017) point out that $b$-dependence of the interstellar $100\,{\rm \mu m}$ intensity largely influences the $b$-dependence of the parameter $b_\lambda$, as well as the anisotropic scattering by interstellar dust grains.
According to their analysis,  the $b_\lambda$ value can change by $\sim \pm 20\%$ in high-$b$ regions in the near-IR.  
Since high-$b$ $100\,{\rm \mu m}$ intensity is an order of $1\,{\rm MJy\,sr^{-1}}$ in diffuse interstellar medium, the uncertainty is estimated as $20\%$ of the DGL intensity when $I_{100,i} = 1\,{\rm MJy\,sr^{-1}}$.
The values are listed as ``DGL $b$-dependence'' in Table 5.

In Section 6.2, we assume the SED of the isotropic IPD to be that of the Kelsall model, but the spheroidal model fitted to the $\epsilon$-dependence at $12\,{\rm \mu m}$ predicts the different SED against the Kelsall model (Fig. 18).
To take into account the SED uncertainty of the isotropic IPD, the $C_\lambda$ difference between the spheroidal model and Model B (Table 4) is assumed as an order of the SED uncertainty for both Model A and B.
These values are listed as ``ZL SED'' in Table 5.
The albedo or phase function of the isotropic IPD may be different from those of the Kelsall model in addition to the density distribution.
However, we assume that the albedo or phase function of the isotropic IPD are similar to those of the Kelsall model because both dust components are thought to be overlapped partly around the sun.

In each case of Model A or B, the total uncertainties of the EBL intensity are calculated as sum of the uncertainty components listed in Table 5.
In Table 5, the EBL results in Model A and B are denoted as ``EBL (Model A)'' and ``EBL (Model B)'', respectively.
According to their results, the EBL detection more than $\sim3\sigma$ is achieved at the shorter near-IR wavelengths, $1.25$, $2.2$, and $3.5\,{\rm \mu m}$ in Model B. 
Therefore, we discuss the EBL at these three wavelengths in the following section.

At $3.5$ and $4.9\,{\rm \mu m}$, the $\epsilon$-dependence of Model A is larger than that of Model B, while both models show the similar $\epsilon$-dependence in the other wavelengths (Fig. 20).
The $\epsilon$-dependence of Model A seems to be caused by the relatively large values of $a_\lambda$, $1.153\pm0.028$ and $1.100\pm0.051$ at $3.5$ and $4.9\,{\rm \mu m}$, respectively (Table 2).
These values are derived by the fitting of the Kelsall model to the DIRBE $\epsilon=90^\circ$ maps (Sano et al. 2016a).
If we require that the EBL should be isotropic as the isotropy test of Hauser et al. (1998), Model B seems better representation than Model A.

\begin{table*}
\begin{center}
 \renewcommand{\arraystretch}{1.0}
 \caption{Resultant EBL intensity and uncertainties in units of ${\rm nW\,m^{-2}\,sr^{-1}}$}
  \label{symbols}
  \begin{tabular}{lrrrrrrr}
  \hline
   Band ($\rm{\mu m}$) & $1.25$ & $2.2$ & $3.5$ & $4.9$ & $12$ & $25$ & $60$ \\
   \hline
    \hline
  Model A ($a_\lambda > 1$) & &  & & & & & \\
 \hline
   Residuals                                                              & $64.23$ & $28.79$ & $8.99$ & $4.55$ & $88.73$ & $96.68$ & $24.26$\\
   Isotropic IPD ($C_\lambda a_\lambda f(\epsilon)$)   & $8.01$ & $3.12$ & $1.75$ & $7.03$ & $87.95$ & $75.19$ & $8.77$\\ 
   Residuals -  Isotropic IPD                                         & $56.22$ & $25.67$ & $7.24$ & $-2.48$ & $0.79$ & $21.49$ & $15.49$\\
   \hline
   Absolute calibration      & $1.99$ & $0.89$ & $0.28$ & $0.14$ & $4.53$ & $14.60$ & $2.52$\\
   $\epsilon$-dependence (upper/lower) & $9.80/5.02$ & $1.81/1.98$ & $1.25/3.68$ & $5.85/12.25$ & $10.52/7.62$ & $14.07/17.99$ & $3.62/7.69$\\
   \hline
    \hline
   Model B ($a_\lambda=1$)& &  & & & & & \\
 \hline
 Residuals                                                                & $64.23$ & $28.79$ & $ 18.59$ & $30.18$ & $206.51$ & $192.80$ & $29.50$\\   
 Isotropic IPD ($C_\lambda a_\lambda f(\epsilon)$)     & $19.45$ & $7.57$ & $3.68$ & $15.52$ & $206.12$ & $176.48$ & $20.97$\\     
   Residuals -  Isotropic IPD                                         & $44.78$ & $21.22$ & $ 14.90$ & $14.66$ & $0.39$ & $16.32$ & $8.53$\\
   \hline
   Absolute calibration       & $1.99$ & $0.89$ & $0.58$ & $0.91$ & $10.53$ & $29.11$ & $3.07$\\
   $\epsilon$-dependence (upper/lower) & $7.44/4.43$ & $0.89/1.75$ & $1.20/0.89$ & $1.65/1.60$ & $13.25/10.12$ & $14.99/11.93$ & $3.62/6.37$\\
   \hline
   \hline
   Common uncertainties& &  & & & & & \\
   \hline
      Statistical                   & $0.06$ & $0.03$ & $0.02$ & $0.06$ & $0.47$ & $0.53$ & $0.09$\\
   DGL $b$-dependence    & $0.94$ & $0.29$ & $0.24$ & $0.17$ & $2.31$ & $1.15$ & $1.71$\\
   ZL SED    & $0.36$ & $1.12$ & $1.35$ & $2.20$ & --- & $8.20$ & $1.08$\\
      \hline  
          \hline
          EBL (Model A) & $56_{-8}^{+13}$ & $26\pm4$ & $7_{-6}^{+3}$ &  $-2_{-15}^{+8}$ & --- & $21_{-42}^{+39}$ & $16_{-13}^{+9}$\\
          EBL (Model B) & $45_{-8}^{+11}$ & $21_{-4}^{+3}$ & $15\pm3$ &  $15\pm5$ & --- & $16_{-51}^{+54}$ & $9_{-12}^{+10}$\\
          \hline
        \end{tabular}
    \end{center}
    \medskip
 \end{table*}

\subsection{Potential isotropic components in our solar system or Galaxy}

We expect the intensity of the scattered light and thermal emission from the isotropic IPD of the filling structure (e.g., OCC dust) to show the $\epsilon$-dependence (Section 2.2).
Though such a component is thought to be removed by the fitting to the $\epsilon$-dependence (Section 6.2), other isotropic components in the outer solar system or Galaxy do not show the $\epsilon$-dependence and could influence the resultant EBL intensity.
Here we discuss the potential contribution of such components. 

In the far-IR wavelengths $\lambda \sim 100\,{\rm \mu m}$, several observations report the EBL intensity higher than the IGL brightness (Odegard et al. 2007; Lagache et al. 2000; Dole et al. 2006; Berta et al. 2011; Matsuura et al. 2011; B\'ethermin et al. 2012).
Tsumura (2018) assume that the intensity difference originates from thermal emission from dust shells in the outer solar system ($> 200\,{\rm AU}$) and estimate the corresponding mass of such a component.
They expect the intensity of the scattered light from the assumed dust shell to be less than $1\,{\rm nW\,m^{-2}\,sr^{-1}}$ at $\lambda \sim1\,{\rm \mu m}$.
Since this is far below the EBL intensity reported so far, as well as the present study (Table 5), we can neglect the contribution from the dust shell.

An influx of interstellar dust to our solar system has been reported by dust detectors onboard spacecrafts, such as {\it Ulysses} and {\it Cassini} (Gr\"un et al. 1993; Gr\"un et al. 1994; Grogan et al. 1996; Altobelli et al. 2016).
Rowan-Robinson \& May (2013) and Kondo et al. (2016) develop the ZL models including an isotropic component on the basis of the mid-IR observations.
They interpret the derived isotropic component as the thermal emission from interstellar dust in our solar system.
However, the mass flux of such a component is reportedly lower than the IPD grains by several orders of magnitude (Gr\"un et al. 1994).
In addition, it would be reasonable to assume that the interstellar dust shows the filling structure, similar to the OCC dust, because they are likely to exist throughout the solar system with some anisotropy.
Therefore, we assume negligible contribution from the interstellar dust.

To interpret the far-IR isotropic residuals observed with DIRBE, Dwek et al. (1998) investigate the possibility of a dust shell surrounding the Milky Way, which can be a potential source of the isotropic emission.
To explain the far-IR residual intensity, they expect the mass of such a component to be as large as that of Galactic disk and conclude that the isotropic emission is not likely from the Galactic component.
Therefore, we also neglect the contribution from such a component to the near-IR EBL. 

\section{IMPLICATION FROM THE PRESENT RESULT OF THE NEAR-IR EBL}

\subsection{Discussion on origin of the EBL excess}

Figure 21 shows the near-IR EBL intensity derived in the present study (Model B) in comparison with the other studies of the EBL and IGL.
In the visible and near-IR wavelengths, the IGL results from deep galaxy observations (Madau \& Pozzetti 2000; Totani et al. 2001; Fazio et al. 2004) are marginally consistent with several   IGL models created by assuming the redshift evolution of galaxies (Dom\'inguez et al. 2011, Inoue et al. 2013, Haadt \& Madau 2012, Helgason \& Kashlinsky 2012, Finke et al. 2010, Gilmore et al. 2012).
Other IGL models are presented by, e.g., Nagamine et al. 2006, Stecker et al. 2006, Stecker et al. 2016, Kneiske \& Dole 2010, and Dwek \& Krennrich (2013) . 
Our result implies that intensity of the EBL at $1.25$ and $2.2\,{\rm \mu m}$ is more than twice as high as that of the IGL, even with the evaluation of the isotropic IPD component.

In the visible wavelengths, some EBL measurements independent of the ZL subtraction are shown in Fig. 21.
Mattila et al.  (2017ab) perform the ``dark cloud method'', which utilizes the attenuation of the EBL at a dense dark cloud in our Galaxy and regard the intensity difference between the surrounding field and cloud as the EBL.
They report the EBL twice as large as the IGL at $0.4\,{\rm \mu m}$.
Zemcov et al. (2017) provide an upper limit of the EBL by analyzing the data obtained with the Long Range Reconnaissance Imager (LORRI) onboard {\it New Horizons}.
The data were taken during the cruising phase toward the outer solar system at $R>5\,{\rm AU}$, where the ZL intensity is expected to be lower than that in the earth orbit by a few orders of magnitude (e.g., Fig. 4 of Zemcov et al. 2018).
Matsumoto et al. (2018) present  reanalysis of the data obtained with a visible camera onboard {\it Pioneer 10/11} beyond $R=3\,{\rm AU}$ (Matsuoka et al. 2011).
They claim presence of an instrumental offset in the data and doubt the result of Matsuoka et al. (2011), who derived low residual intensity comparable to the IGL. 
Matsumoto \&  Tsumura (2019) give a lower limit of the EBL according to their analysis of auto- and cross-correlations of visible images of {\it Hubble} Extreme Deep Field (XDF; Illingworth et al. 2013).
These upper and lower limits are about twice as large as the IGL.
These results suggest room for extragalactic components other than normal galaxies in the visible wavelengths.
Combined with these limits and present result, the EBL is thought to have a spectrum with its peak intensity at $\lambda \sim 1\,{\rm \mu m}$.

A number of theoretical studies investigate potential contribution of first stellar objects at $z \gtrsim 6$, such as primordial stars and blackholes, to the EBL since the UV radiation is redshifted to the visible and near-IR wavelengths in the present epoch (e.g., Cooray \& Yoshida 2004; Dwek et al. 2005a; Cooray et al. 2012a; Fernandez \&  Komatsu 2006; Madau \& Silk 2005; Mii \&  Totani 2005; Salvaterra \& Ferrara 2003; Santos et al. 2002; Inoue et al. 2013; Yue et al. 2013).
However, most of the studies predict only small contribution of such sources to the EBL.
Moreover, the UV radiation from the early universe is expected to have a spectral cutoff in the shortest wavelength due to the absorption by neutral hydrogen in the intergalactic medium.
Since we do not clearly see the sharp edge at $\lambda \sim 1\,{\rm \mu m}$ in the EBL spectrum (Fig. 21), it would be difficult to assume that the entire excess of the EBL originates from the objects in the early universe.   

Potential contribution from other extragalactic sources at lower redshift has been studied as well.
Cooray et al. (2012b) develop a model of intra halo light (IHL) whose origin is stars tidally stripped out of galaxies during the merger phase.   
Schleicher et al. (2009) and Maurer et al. (2012) assume the dark stars powered by self-annihilating dark matter and the contribution to the EBL intensity.
Mapelli \& Ferrara (2005) calculate contribution from photon created by sterile neutrino decay that can contribute to the EBL.
Though combination of these components can explain the excess in a part, it may be difficult to explain the high EBL intensity at $1.25\,{\rm \mu m}$ as discussed in Paper I.
   
Since the Kelsall ZL model leaves large residuals in the DIRBE map at $25\,\rm{\mu m}$, Wright (1997) defines ``very strong no zodi'' condition that reduces the $25\,\rm{\mu m}$ residuals by a factor of $7$.
Wright (1998) and Gorjian et al. (2000) adopt this condition and develop a parameterized ZL model independent from the Kelsall model, hereafter referred to as the Wright model.
Therefore, it is reasonable that the residuals obtained by the Wright model tend to be smaller than those by the Kelsall model (Fig. 1) .
Arendt \&  Dwek (2003) summarize the near-IR residual intensity derived by using both Kelsall and Wright models.
The present analysis to set the mid-IR residuals to be zero is similar to the idea of ``very strong no zodi'' condition of Wright (1997).
For comparison with our EBL results by the Kelsall model, residual intensity derived with the Wright model is also shown in Fig. 21 (Wright 2001; Levenson et al. 2007).
Their results are marginally consistent with ours, while their intensity tends to be slightly lower than ours at $1.25\,\rm{\mu m}$.
This discrepancy may be caused by the different IPD parameters in the scattering component between the Kelsall and Wright model.
To reduce our EBL intensity at $1.25\,\rm{\mu m}$ to the residual level derived by Levenson et al. (2007), the SED of the ZL from the isotropic IPD in the present analysis should be bluer than the Kelsall model at $\lambda \sim1$--$2\,\rm{\mu m}$, as inferred from Fig. 18.
As for the scattering phase function, the intensity ratio of the scattered light at $1.25\,\rm{\mu m}$ to the  mid-IR thermal emission toward intermediate $\epsilon$ regions ($\epsilon\sim90^\circ$) is expected to increase if the phase function is flatter than that of the Kelsall model (Fig. 16a).
If the flatter phase function is adopted for the  isotropic IPD component, the EBL intensity at $1.25\,\rm{\mu m}$ would decrease and approach the residuals derived by the Wright model.
To reveal the detailed spectrum of the ZL from the isotropic IPD in the near-IR, it is necessary to observe it separately from the smooth cloud component and it is beyond the scope of this paper.

The present study evaluates the isotropic IPD component and derives the high EBL intensity in the near-IR.
This result serves evidence of significant contribution from extragalactic sources other than usual galaxies. 
However, origin of the excess is still unclear due to the large uncertainty associated the ZL evaluation and the wide-band photometric observation with DIRBE, which are insufficient to detect spectral features expected in the theoretical extragalactic sources.

\begin{figure*}
\begin{center}
 \includegraphics[scale=0.8]{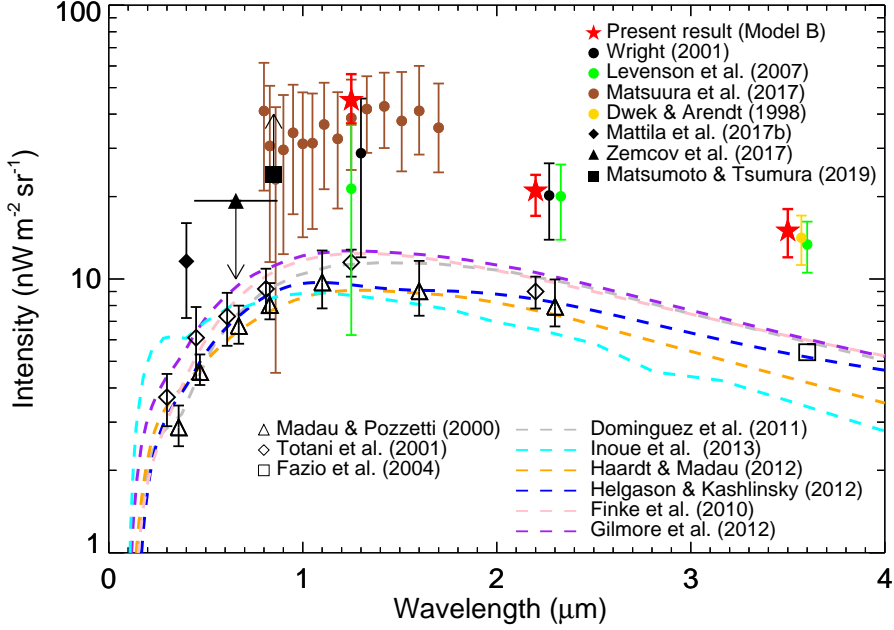}
 \caption
 {Present result of the near-IR EBL intensity in comparison with other studies.
 Our results with the evaluation of the isotropic IPD (Model B) are indicated by red stars (Table 5).
 Visible measurements independent of the ZL subtraction are represented by a filled triangle (Zemcov et al. 2017), diamond (Mattila et al. 2017b), and square (Matsumoto \& Tsumura 2019).
The IGL observations are indicated by the same open symbols as Fig. 1 (Totani et al. 2001; Madau \& Pozzetti 2000;  Fazio et al. 2004).
 Gray, cyan, orange, blue, pink, and purple dashed curves indicate the IGL models developed by Dom\'inguez et al. (2011), Inoue et al. (2013), Haadt \& Madau (2012), Helgason \& Kashlinsky (2012), Finke et al. (2010), and Gilmore et al. (2012), respectively.
 Black and green circles denote the residual intensity derived by Wright (2001) and Levenson et al. (2007), respectively, by using the Wright ZL model (e.g., Wright et al. 1998). 
At $3.5\,\rm{\mu m}$, a gold circle indicates the EBL brightness expected from the present EBL intenisty at $2.2\,\rm{\mu m}$ by assuming the EBL relation between $2.2$ and $3.5\,\rm{\mu m}$ (Dwek \& Arendt 1998).
Brown circles indicate the residual intensity obtained with the {\it CIBER} observation (Matsuura et al. 2017).}
\end{center}
\end{figure*}

\subsection{Spatial fluctuation of the EBL}

To investigate the origin of the near-IR  EBL, spatial fluctuation of the EBL has also been investigated robustly, in parallel with the absolute brightness measurements.
Owing to the large-scale uniformity of the ZL (e.g., Pyo et al. 2012), the fluctuation analysis is useful for the EBL study, free from the ZL contamination.
Analyzing the near-IR data obtained with {\it Spitzer}, {\it AKARI}, and {\it IRTS}, a number of studies claim spatial fluctuation larger than the prediction from clustering of normal galaxies (Cooray et al. 2004; Cooray et al. 2007; Kashlinsky et al. 2004; Matsumoto et al. 2011; Kashlinsky et al. 2005; Kashlinsky et al. 2012; Helgason et al. 2012; Chary et al. 2008; Kim et al. 2019).
Analyzing the data obtained with the {\it CIBER} imager, Zemcov et al. (2014) find large fluctuation of the EBL at $1.1$ and $1.6\,\rm{\mu m}$, and explain it by the IHL model developed by Cooray et al. (2012b).
Several studies report significant coherence between the near-IR and X-ray EBL (e.g., Cappelluti et al. 2017; Helgason et al. 2014; Kashlinsky 2016; Kashlinsky et al. 2019).
These results may suggest that the large fluctuation of the near-IR EBL originates from X-ray sources, such as primordial or direct collapse blackholes.

Analyzing the images of {\it HST} XDF, Matsumoto \& Tsumura (2019) find large fluctuation in the visible four bands. 
As one candidate of the fluctuation, they suggest significant contribution of what they call faint compact objects (FCOs) found in a source catalog of {\it Hubble} Ultra Deep Field.
The number counts of the FCOs increase continuously to fainter end of the catalog, $\sim30{\rm th}$ magnitude.
They estimate that the FCOs can explain the high intensity of the visible  EBL at $\lambda \sim0.8\,\rm{\mu m}$ if the number counts continue to increase up to $\sim35{\rm th}$ magnitude.  
Though the model of the FCOs could explain the excess of the absolute intensity and fluctuation of the EBL simultaneously, it is not necessary to explain their excess by the same sources.
It is possible that the origins of the intensity excess is different from those of the fluctuation.

\subsection{Constraints on the EBL intensity from high-energy $\gamma$-ray observations}

The EBL is known to have cross sections with high-energy photons of $\sim{\rm GeV}$--${\rm TeV}$ via the electron-positron pair creation (Jauch \& Rohrlich 1955). 
Therefore, a number of studies have measured the EBL intensity by observing the spectral attenuation of $\gamma$-rays from blazers in comparison to the assumed intrinsic spectra.
Such observations have been done by High Energy Stereoscopic System (H.E.S.S), MAGIC, and {\it Fermi} (e.g., Dwek et al. 2005b; Schroedter; Orr et al. 2011; Abdollahi et al. 2018; Abramowski et al. 2013; Aharonian et al. 2006; Albert et al. 2008; Abdo et al. 2010; Biteau \&  Williams 2015; Biasuzzi et al. 2019; Korochkin et al. 2020). 
Though a main uncertainty of this method lies in the assumption of the intrinsic spectra of high-energy sources, most of these studies estimate low EBL intensity at most twice as high as the IGL level in the visible and near-IR wavelengths.
Therefore, the $\gamma$-ray constraints on the EBL conflict with the direct EBL observations conducted in the previous and present studies, particularly at $\lambda \sim 1\,\rm{\mu m}$.
This discrepancy has long been controversial in the EBL study.

In the redshift range of $z \gtrsim 0.2$--$3$, the {\it Fermi}-LAT Collaboration shows consistency between the $\gamma$-ray observations and galaxy evolution models from which the IGL intensity is calculated (Fig. 1 of Abdollahi et al. 2018).
This result suggests little room for the excess component of the EBL at $z \gtrsim 0.2$--$3$, but it could exist at low redshift of $z \lesssim 0.1$.
This may indicate that additional extragalactic sources, such as the FCOs, can explain the EBL excess if they exist at $z \lesssim 0.1$.
See Matsumoto \&  Tsumura (2019) for more quantitative discussion on the FCOs.

Due to the tension between the $\gamma$-ray observation and residual intensity, particularly derived by Low Resolution Spectrometer (LRS) of {\it CIBER} (Matsuura et al. 2017),  Kohri \& Kodama (2017) investigate a possible mixing between the EBL photons and axions to increase the transparency for the $\gamma$-ray.
Adopting this theory to the {\it CIBER} result, they constrain parameters of the axion mass and axion-photon coupling constant that can solve the friction between the EBL and $\gamma$-ray observations.  
Since the present result of the EBL at $1.25\,\rm{\mu m}$ is consistent with the {\it CIBER} residuals (Fig. 21), our result can avoid the conflict with the $\gamma$-ray observations as well, by assuming the coupling of the EBL photons and axions.

\subsection{Future prospect of the EBL observation}

With the quantitative evaluation of the isotropic IPD component, the present study helps to consolidate the idea of the high intensity of the near-IR EBL in comparison with the IGL.
However, the origin of the excess cannot be identified in the present study.
To reveal the EBL origin, further observations are necessary.

Simultaneous observations of the EBL intensity and fluctuation in visible and near-IR wavelengths will be useful to probe the origin of the EBL.
A new sounding rocket project, {\it Cosmic Infrared Background Experiment  2} ({\it CIBER-2}), is designed to conduct both imaging and spectrometry in the wavelengths of $0.5$--$2.0\,\rm{\mu m}$.
With a large telescope of a $28.5\,\rm{cm}$ diameter, {\it CIBER-2} will achieve $10$ times more sensitivity than {\it CIBER} for diffuse radiation of our interest (Lanz et al. 2014; Shirahata et al. 2016; Nguyen et al. 2018).
In addition, {\it CIBER-2} has large field of view of $\sim2.3^\circ \times 2.3^\circ$ and high spectral resolution for the diffuse light measurement ($\lambda/\Delta\lambda \sim 20$).
{\it CIBER-2} plans to launch in 2020 in cooperation with international collaborators and NASA Sounding Rocket Operations Contract (NSROC).

A future project, {\it The Spectro-Photometer for the History of the Universe, Epoch of Reionization and Ices Explorer} ({\it SPHEREx}) is NASA's mid-class satellite mission and plans to launch in 2023.
{\it SPHEREx} will carry out the first all-sky spectral survey at $\sim0.75$--$5.0\,\rm{\mu m}$, covering the near-IR wavelengths longer than {\it CIBER-2}. 
Since the sensitivity of {\it SPHEREx} will be higher than that of {\it CIBER} by more than two orders of magnitude, {\it SPHEREx} is capable of measuring the large-scale fluctuation of the EBL expected to originate from the epoch of reionization ($z \gtrsim 6$).
Moreover, the all-sky spectral observations are useful to construct a new ZL model with higher spectral resolution  than the previous ones.

In addition to the precise observations from the earth orbit, it will be extremely beneficial to observe the sky from deep space ($R \gtrsim 5\,{\rm AU}$), where the ZL intensity is expected to be lower than that around the earth by more than one order of magnitude (Zemcov et al. 2018).
In the visible wavelengths, such an opportunity has been provided by instruments onboard spacecrafts, such as {\it Pioneer 10/11} (Matsumoto et al. 2018) and {\it New Horizons} (Zemcov et al. 2017).
Targeting opportunities of the future spacecrafts cruising beyond the Jupiter orbit ($R \gtrsim 5\,{\rm AU}$), we have been developing a visible and near-IR spectroscopic instrument, what we call Exo-Zodiacal Infrared Telescope (EXZIT; Matsuura et al. 2014).
Observation with EXZIT will allow us to confirm the contribution of the isotropic IPD component and the EBL intensity.
The deviation of the observed isotropic IPD from the simple model (Fig. 14) may imply that the density structure of the isotropic IPD is different from the prediction of $\sim 1/R$ (Section 5.1).
The deep-space observations with EXZIT will be useful to probe the structure of the isotropic IPD component thanks to the promising decrease of the main IPD component from JFCs (Zemcov et al. 2018).

\section{SUMMARY}

We present the study on the isotropic IPD component and EBL on the basis of the IR observations with DIRBE.
Since the intensity of the scattered light and thermal emission from the isotropic IPD is expected to show the $\epsilon$-dependence, we investigate that trend by the DIRBE weekly-averaged maps at $1.25$, $2.2$, $3.5$, $4.9$, $12$, $25$, and $60\,{\rm \mu m}$, which cover the wide $\epsilon$ range of $64^\circ \lesssim \epsilon \lesssim 124^\circ$.
After subtracting the other emission components, the Keslall ZL model, ISL, and DGL, from the DIRBE intensity maps, we investigate the residuals as a function of $\epsilon$.
We find the $\epsilon$-dependence of the residual intensity at each wavelength, indicating the presence of the isotropic IPD that is not included in the Kelsall model.
However, the observed $\epsilon$-dependence shows the deviation from the simple model of the isotropic IPD in the high-$\epsilon$ regions.
The $\epsilon$-dependence could be explained by assuming the spheroidal cloud of higher IPD density further away from the sun.
In addition, the discrepancy could be contributed from the uncertainties of the density distribution, phase function, or $R$-dependence of the IPD temperature assumed in the Kelsall model.
According to the residual level in the mid-IR, the density of the isotropic IPD is estimated to be $\sim5\%$ of that of the total IPD, consistent with the earlier studies.

To evaluate the EBL intensity from the derived residuals, we fit the $\epsilon$-dependence of the residuals at $12\,{\rm \mu m}$ by a polynomial function and expect the intensity of the isotropic IPD in the near-IR by assuming the SED of the ZL.
As the result of the separation of the EBL from the isotropic IPD, the intensity of the EBL is $45_{-8}^{+11}$, $21_{-4}^{+3}$, and $15\pm3\,{\rm nWm^{-2}sr^{-1}}$ at $1.25$, $2.2$, and $3.5\,{\rm \mu m}$, respectively.
The EBL intensity at $1.25$ and $2.2\,{\rm \mu m}$ is a few times higher than the IGL, indicating that the additional extragalactic sources are the predominant emission component in comparison with normal galaxies, although the origin of the excess remains unclear.
The high intensity of the near-IR  EBL could avoid the friction with the measurements from high-energy $\gamma$-ray observations if the origin of the excess is present in low redshift or hypothetical process of the photon-axion mixing increases the transparency for the $\gamma$-rays.

\acknowledgments

We are grateful to the anonymous referee for a number of constructive and insightful comments to improve the manuscript.
We acknowledge a number of useful discussions with Takafumi Ootsubo, Daisuke Yonetoku, Kohji Tsumura, and Kenta Dambayashi.
K.S. is supported by Grant-in-Aid for Japan Society for the Promotion of Science (JSPS) Fellows.
This work was supported by JSPS KAKENHI Grant Number 15H05744 and 19J00468. 






\appendix

\clearpage



\clearpage









\clearpage

\end{document}